\documentclass{aastex61}

\usepackage{natbib}

\def\solum{$L_\odot$}
\def\solmass{$M_\odot$}
\def\lamb{$\lambda$}
\def\lax{{$\mathrel{\hbox{\rlap{\hbox{\lower4pt\hbox{$\sim$}}}\hbox{$<$}}}$}}
\def\gax{{$\mathrel{\hbox{\rlap{\hbox{\lower4pt\hbox{$\sim$}}}\hbox{$>$}}}$}}

\def\nd{\nodata}
\newdimen\digitwidth      
\setbox1=\hbox{0}       
\digitwidth=\wd1        
\catcode`"=\active      
\def"{\kern\digitwidth}

\newcommand{\mbh}{\ensuremath{M_\mathrm{BH}}}
\newcommand{\msigma}{\ensuremath{M_{\mathrm{BH}}-\sigs}}
\newcommand{\hii}{\ion{H}{2}}
\newcommand{\oii}{[\ion{O}{2}]}
\newcommand{\oiii}{[\ion{O}{3}]}
\newcommand{\sigs}{\ensuremath{\sigma_{\ast}}}
\newcommand{\sigg}{\ensuremath{\sigma_g}}
\newcommand{\siggc}{\ensuremath{\sigma^c_g}}
\newcommand{\siggs}{\ensuremath{\sigma^s_g}}
\newcommand{\kms}{${\rm km~ s^{-1}}$}

\newcommand{\sii}{[\ion{S}{2}]}
  
\newcommand{\delsigc}{$\Delta \sigma^c$}  
\newcommand{\nii}{[\ion{N}{2}]}

\newcommand{\lum}{erg s$^{-1}$}

\newcommand{\lledd}{${\it L_{\rm{bol}}/L{\rm{_{Edd}}}}$}
\newcommand{\loiii}{\ensuremath{L_{\mathrm{[O~ { III}]}}}}

\newcommand{\mgb}{\ion{Mg}{1}{$b$}}

\newcommand{\sn}{S/N}

\newcommand{\cahk}{\rm Ca H+K}

\newcommand{\Ne}{[\ion{Ne}{3}] $\lambda 3968$}

\def\kms{km s$^{-1}$}

\newcommand{\sigc}{\ensuremath{\sigma^c_g}}

\newcommand{\hbeta}{H\ensuremath{\beta}}
\newcommand{\halpha}{H\ensuremath{\alpha}}

\def\lax{{$\mathrel{\hbox{\rlap{\hbox{\lower4pt\hbox{$\sim$}}}\hbox{$<$}}}$}}
\def\gax{{$\mathrel{\hbox{\rlap{\hbox{\lower4pt\hbox{$\sim$}}}\hbox{$>$}}}$}}

\shorttitle{Black Hole Masses and Eddington Ratios of Type 2 Quasars}
\shortauthors{Kong \& Ho}

\begin{document}

\title{The Black Hole Masses and Eddington Ratios of Type 2 Quasars}
\correspondingauthor{Luis C. Ho}
\email{lho.pku@gmail.com}

\author{M. Z. Kong}
\affil{Department of Physics, Hebei Normal University, No.20 East of South 2nd Ring Road, Shijiazhuang 050024, China}

\author{Luis C. Ho}
\affil{Kavli Institute for Astronomy and Astrophysics, Peking University,
Beijing 100871, China}
\affil{Department of Astronomy, School of Physics, Peking University,
Beijing 100871, China}

\begin{abstract}
Type~2 quasars are an important constituent of active galaxies, possibly 
representing the evolutionary precursors of traditionally studied type~1 
quasars.  We characterize the black hole mass (\mbh) and Eddington ratio 
(\lledd) for 669 type~2 quasars selected from the Sloan Digital Sky Survey, 
using black hole masses estimated from the \msigma\ relation and bolometric 
corrections scaled from the extinction-corrected \oiii\ \lamb 5007 luminosity.
When stellar velocity dispersions cannot be measured directly from the spectra,
we estimate them from the core velocity dispersions of the narrow emission 
lines \oii\ \lamb\lamb 3726, 3729, \sii\ \lamb\lamb 6716, 6731, and \oiii\ 
\lamb 5007, which are shown to trace the gravitational potential of the stars.
Energy input from the active nucleus still imparts significant perturbations to
the gas kinematics, especially to high-velocity, blueshifted wings.  Nonvirial 
motions in the gas become most noticeable in systems with high Eddington ratios.
The black hole masses of our sample of type~2 quasars range from $M_{\rm BH}
 \approx 10^{6.5} $ to $10^{10.4} \, M_\odot$ (median $10^{8.2} \, M_\odot$).
Type~2 quasars have characteristically large Eddington ratios (\lledd\ 
$\approx 10^{-2.9} - 10^{1.8}$; median $10^{-0.7}$), slightly higher than in 
type~1 quasars of similar redshift; the luminosities of $\sim$20\% of the 
sample formally exceed the Eddington limit.  The high Eddington ratios may be 
consistent with the notion that obscured quasars evolve into unobscured 
quasars.  
\end{abstract}

\keywords{galaxies: active --- galaxies: nuclei --- galaxies: Seyfert --- 
quasars: emission lines }

\section{Introduction}

The classical unified model of active galactic nuclei (AGNs) has served as a 
useful framework for synthesizing a wide range of observations (Antonucci 
1993; Urry \& Padovani 1995; Netzer 2015).  In the simplest version of this 
model, the orientation of the central engine with respect to the viewer plays 
a major role in determining the observed characteristics, and hence 
classification, of AGNs.  In particular, obscuration along the 
line-of-sight by a parsec-scale dusty torus dictates whether or not emission 
from the broad-line region is directly visible: unobscured type~1 sources 
exhibit broad permitted lines with velocity widths [full width at half maximum 
(FWHM)] \gax\ 1000 \kms, whereas obscured type~2 sources display only narrow 
emission lines.  Within this framework, both types are intrinsically the same.  

In recent years, however, it has become increasingly clear that the 
conventional, purely orientation-based unified model cannot account for the
full diversity of the AGN population.  Nor is it reasonable that it should.  
Accretion onto a central supermassive black hole (BH) powers AGNs. To the 
extent that the general galaxy population certainly evolves strongly with 
cosmic epoch, so, too, must the population of active galaxies, as BHs coevolve 
with their hosts (Kormendy \& Ho 2013).  The regime of low mass accretion rates
(Ho 2008) offers the most dramatic examples of departures from the ``standard''
paradigm.  At the lowest accretion rates, the obscuring torus---indeed, even 
the broad-line region itself---disappears (Elitzur \& Ho 2009; Elitzur 
et al. 2014).  Although less well-studied, ``intrinsic'' type~2 AGNs also 
emerge under conditions of higher accretion rates (e.g., Ho et al. 
2012; Miniutti et al. 2013; Elitzur \& Netzer 2016; Bianchi et al. 2017; 
Shu et al. 2017).  Furthermore, in the merger-driven picture of quasar 
evolution (Sanders et al. 1988; Hopkins et al. 2006), dust-obscured (type~2) 
AGNs arise not simply from geometric viewing angle effects but instead as a 
direct consequence of a particular evolutionary phase, characterized by 
different, physically distinct conditions on both large and small scales.  In 
other words, in the evolution-based model type~1 and type~2 AGNs are
{\it not}\ intrinsically the same.

The evolutionary scenario linking type~1 and type~2 AGNs should manifest itself 
most readily in sources of high luminosity.  During the past 15 years, 
increasing attention has been drawn to a population of type~2 quasars, 
especially the large samples uncovered systematically at optical
wavelengths from the Sloan Digital Sky Survey (SDSS).  Zakamska et al. (2003) 
identified 291 objects with redshifts $0.037 < z < 0.83$ based on standard 
optical diagnostic emission-line intensity ratios that place them in the 
category of AGNs.  These objects have sufficiently narrow (FWHM $< 2000$ \kms) 
permitted lines that make them likely candidates for type~2 sources, and, at 
the same time, about 50\% of them have high enough \oiii\ $\lambda 5007$ 
luminosities ($> 3\times 10^{8}\ L_{\odot}$) that, when translated to 
equivalent $B$-band absolute magnitudes, qualify them as quasars by the 
historical criterion of $M_B < -23$ mag (Schmidt \& Green 1983).  Following a 
similar method, Reyes 
et al. (2008) expanded the SDSS sample to 887 candidates, the largest catalog 
to date of optically selected low-to-moderate redshift type~2 quasars\footnote{
Reyes et al. (2008) adopt a slightly modified luminosity criterion of \loiii\ 
$> 2\times 10^{8}\ L_{\odot}$.}.  More recently, Alexandroff et al. (2013) 
further used SDSS to extend the selection of type~2 quasars to $2 < z < 4.3$.

What are the basic properties of type~2 quasars, and how do they compare with 
those of the more traditionally studied type~1 quasars?  A number of studies 
suggest that the host galaxies of the two types of quasars indeed appear to 
differ.  Relative to their type~1 counterparts, type~2 quasars seem to have 
elevated star formation rates (Kim et al. 2006; Zakamska et al. 2008, 2016), a 
higher incidence of tidal disturbances (Bessiere et al. 2012), and lower halo 
masses (Ballantyne 2016).  Type 2 quasars also have a higher frequency of 
flat-spectrum radio cores (Lal \& Ho 2010).  

BH mass is one of the most fundamental physical parameters for AGNs.  Also of 
interest for understanding the evolutionary state and the interconnections 
between different AGN populations is their Eddington ratio\footnote{The 
Eddington ratio is defined as \lledd, with $L_{\rm bol}$ the bolometric 
luminosity and $L_{\rm Edd}=1.26\times10^{38}~(M_{\rm BH}/M_{\rm \odot})
~\rm erg~s^{-1}$, with $M_{\rm BH}$ the mass of the BH.}, which, too, 
relies on knowledge of the BH mass.  Significant progress 
has been made in developing methods to derive BH masses for type~1 AGNs, 
through either reverberation mapping of broad emission lines (e.g., Kaspi et 
al. 2000) or empirically calibrated formalisms that use single-epoch spectra
(e.g., Greene \& Ho 2005b; Vestergaard \& Peterson 2006; Ho \& Kim 2015).  
These techniques, however, cannot be applied to narrow-line AGNs.  Apart from 
a few rare instances in which BH masses in active galaxies can be measured 
directly through spatially resolved kinematics (see review in Kormendy \& Ho 
2013), in general we must resort to indirect methods to estimate BH masses for 
type~2 AGNs, principally through the well-known correlations between BH mass 
and host galaxy bulge stellar mass or velocity dispersion.  These empirical 
relations, of course, were established originally for inactive galaxies at 
$z \approx 0$ (e.g., Tremaine et al. 2002; H\"aring \& Rix 2004).
When applying these relations to more active systems and 
especially at higher redshift, we implicitly assume that (1) AGNs obey the 
same empirical scaling relations as local inactive galaxies, (2) these 
relations do not evolve with redshift, or (3) any potential redshift 
dependence can be corrected.

In this paper, we investigate the possibility of estimating BH masses for 
type~2 quasars using the \msigma\ relation (Ferrarese \& Merritt 2000; Gebhardt 
et al. 2000).  We proceed in two steps.  Using the SDSS-selected sample of 
type~2 quasars from Reyes et al. (2008), we directly measure \sigs\ for a 
subset of sources having spectra of sufficiently high quality with detectable 
stellar features.  For the majority of the sample for which these conditions 
could not be met, we estimate
\sigs\ indirectly from the widths of their nebular emission lines.  In AGNs of 
lower luminosity, it has been well-established that the kinematics of the 
ionized gas in the narrow-line region generally trace the virial motions 
of the stars of the host galaxy bulge (e.g., Whittle 1992; Nelson \& Whittle 
1996).  In particular, in Seyfert galaxies the velocity width of \oiii\ \lamb 
5007 traces \sigs, albeit with considerable scatter.  Nelson (2000) found that 
\sigg, the velocity dispersion derived from ionized gas, can be used as a proxy
to study the \msigma\ relation of AGNs.  Greene \& Ho (2005a) investigated the 
relation between gaseous and stellar kinematics for a large sample of 
SDSS-selected type~2 AGNs, showing that \sigg\ correlates best with \sigs\ when
the gas velocities are derived from the core of the \oiii\ line, after removing
its asymmetric wings.  The \sigg$-$\sigs\ relation also holds for the lower 
ionization lines \oii\ \lamb\lamb 3726, 3729 and \sii\ \lamb\lamb 6716, 6731.  
The analysis of Greene \& Ho (2005a) focused on AGNs of relatively low 
luminosity, with \loiii\ \lax\ $10^{8}$ \solum.  Ho (2009) extended these 
results to sources of even lower luminosity, using \nii\ \lamb6583.

Is the \sigg$-$\sigs\ relation applicable to the more powerful type~2 quasars? 
As the AGN luminosity increases, the narrow-line region gas may be increasingly
susceptible to non-gravitational perturbations (e.g., radiation pressure), and 
it is unclear the extent to which the kinematics of the ionized gas still track
the virial velocity of the underlying stellar gravitational potential.  Greene 
et al. (2009) analyzed a sample of 111 type~2 quasars and found that the line 
widths of \oii\ and \oiii\ exhibit no significant correlation with \sigs.  This 
suggests that \sigg\ cannot be used to estimate BH mass for AGNs with 
luminosities in the range of quasars.

We investigate this problem with a new, independent analysis of the 
SDSS-selected sample of type~2 quasars of Reyes et al. (2008), taking special 
care to account for factors that may influence \sigg, such as asymmetries and 
other substructure in the line profile.  Although the scatter is considerable, 
we find that, as in lower-luminosity AGNs, \sigg\ does still trace \sigs\ 
usefully in type~2 quasars.  We use our newly derived \sigg$-$\sigs\ relation, 
in combination with the latest \msigma\ relation, to estimate BH masses and 
Eddington ratios for the Reyes et al. (2008) sample of low-redshift type~2 
quasars.

This paper is organized as follows.  Section 2 discusses the sample selection.
Section 3 presents measurements of \sigs\ and the analysis of the emission 
lines to derive \sigg.  Section 4 compares \sigg\ and \sigs, discusses the 
prevalence of novirial motions in the narrow-line region, and applies the velocity dispersions
to estimate BH masses and Eddington ratios.  We summarize our main conclusions 
in Section 5.  We adopt the following cosmological parameters: $\Omega_m$ = 
0.3, $\Omega_{\Lambda}$ = 0.7, and $H_0$ = 70 $\rm km~ s^{-1} ~Mpc^{-1}$.  
 
\section{Sample}

As one of the main purposes of our work is to measure stellar velocity 
dispersions and compare them with velocity dispersions derived from ionized gas,
we need access to the rest-frame optical stellar continuum and a variety of 
well-measured strong emission lines.  Thus, we focus on the sample of 887 type~2
quasars with redshifts $0.037< z< 0.83$ from Reyes et al. (2008). The relatively
low redshifts of the sample also help to minimize possible redshift evolution 
of the \msigma\ relation, an assumption critical to our application of the 
velocity dispersions to estimate BH masses for the sample.

We obtain the optical spectra of the sample from the seventh data release of 
SDSS (Abazajian et al. 2009).  Some of the objects exhibit obvious 
double-peaked or otherwise conspicuous complex substructure in their 
emission-line core.  The Reyes et al. sample contains a total of 126 sources 
with complex line substructure, among them 42 that have unambiguous 
double-peaked profiles.  The line substructure is most evident in \oiii, 
presumably because of its high signal-to-noise ratio (\sn), but in many 
cases it can also be seen in the weaker low-ionization lines.  The objects 
with complex profile substructure span a similar range of \oiii\ luminosity 
and redshift as the parent sample (Figure 1).  A variety of physical origins 
may give rise to kinematic complexity in the narrow-line gas.  For instance, 
double-peaked profiles may be indicative of galaxy mergers, large-scale disk 
rotation, or outflows (e.g., Liu et al. 2010;  Ge et al. 2012).   The 
interaction of compact radio jets with the interstellar medium of the host 
galaxy can produce flat-topped cores and other profile substructure 
(e.g., Whittle et al. 1988; Gelderman \& Whittle 1994).  Whatever the case may 
be, these complexities will complicate the interpretation of the line widths 
and compromise our ability to use them to trace virial velocities.  We omit 
the 126 sources with complex profiles from our analysis of the emission-line 
widths, but we retain them for the purposes of measuring stellar velocity 
dispersions.

\section{Analysis}

\subsection{Stellar Velocity Dispersions}

\subsubsection{Method}

The basic principle for measuring the line-of-sight stellar velocity 
dispersion of a galaxy through integrated spectroscopy is to estimate the 
degree to which the intrinsic stellar spectrum of the galaxy has been broadened
by the internal motions (velocity dispersion) of its constituent stars.  
Traditional measurements are mostly based on Fourier or cross-correlation 
techniques (e.g., Sargent et al. 1977; Tonry \& Davis 1979; Bender 1990), but 
more modern studies increasingly favor methods based on direct fitting of the 
spectra in pixel space (e.g., Kelson et al. 2000; Barth et al. 2002; 
Cappellari \& Emsellem 2004; Greene \& Ho 2006, hereafter GH06).  One of the 
advantages of the latter method is that it is straightforward to exclude 
emission lines and other unwanted or bad pixels from the fit.  The robustness 
of the fit can be evaluated readily from direct comparison of the 
broadened template spectrum with the galaxy spectrum.  
 
We use the publicly available penalized pixel-fitting code {\tt pPXF} of 
Cappellari \& Emsellem (2004) to measure \sigs.  Prior to analysis, we correct 
the SDSS spectra for foreground dust reddening, using the extinction values 
of Schlafly \& Finkbeiner (2011) and the Galactic extinction curve of Cardelli 
et al. (1989).  The spectra were then shifted to their rest frame, based on 
SDSS-derived redshifts.  Our adopted values for the $V$-band Galactic 
extinction and redshift are listed in Table 1.
 
The {\tt pPXF} program models the galaxy spectrum as

\begin{equation}
M_{\rm mod}(x) = P(x)~ \left\{ \sum_{j=1}^{N} w_{j} \ [T_{j}(x)~\otimes\ G(x)] \right\}  +  C(x),
\end{equation}

\noindent
where $w_{j}$ is the fractional contribution of the $j$th stellar template 
$T_{j}(x)$ to the model flux, $\otimes\ $denotes convolution, $G(x)$ is a 
Gaussian broadening function with dispersion $\sigma_m$, and $P(x)$ is a 
Legendre polynomial that accounts for potential mismatches in shape between 
the galaxy and the templates, which might arise, for instance, from internal 
reddening, mismatches in stellar population, or residual calibration errors 
(e.g., GH06; Liu et al. 2009).  The term $C(x)$ is a Legendre polynomial 
that represents any additional additive contribution, such as a power-law 
continuum from scattered light from the central AGN or a hot continuum from 
young O and B stars.  The best-fit parameters are obtained by minimization of 
$\chi^{2}$ using the nonlinear Levenberg-Marquardt algorithm implemented in 
the IDL package {\tt mpfit}\footnote{\url{http://purl.com/net/mpfit}} 
(Markwardt 2009).

The template $T(x)$ in principle should faithfully match the stellar population
of the galaxy.  However, in practice, it can be reasonably well represented by 
the spectrum of an optimally weighted linear combination of several stars or of
even a well-chosen single star (e.g., Barth et al. 2002; GH06; Ho et al. 2009).
There are many available libraries of stellar templates (e.g., Bruzual \& 
Charlot 2003; Valdes et al. 2004; S$\acute{\rm a}$nchez-Bl$\acute{\rm a}$zquez 
et al. 2006), which have different wavelength coverage and spectra resolution.  
In light of the instrumental spectral resolution of SDSS (mean FWHM = 
$2.94\pm0.31$ \AA\ over the region 4100--5400 \AA\ used for our analysis of 
\sigs; see below), the most suitable choice for our applications is the 
Indo-U.S. stellar spectral library (Valdes et al. 2004), which 
covers the wavelength range 3460--9464 \AA\ at a spectral resolution of FWHM 
= 1.35 \AA\ (Beifiori et al. 2011).  Consistent with prior studies, a 
weighted linear combination of late-type (F, G, K, and M) red giant 
(luminosity class III) stars of near-solar metallicity provides a good match 
to the stellar continuum of the host galaxies.  We also include an A-type 
dwarf (luminosity class V) star to account for those galaxies that have a 
substantial post-starburst stellar population.  

As the template stars used in our fits were observed with a substantially 
higher spectral resolution than the SDSS science targets, the model velocity 
dispersion ($\sigma_m$) needs to be corrected to yield the intrinsic stellar 
velocity dispersion (\sigs) of the galaxy.  We adopt $\sigs^2 = \sigma_{m}^2
+ \sigma_{\rm inst, V}^2 - (\sigma_{\rm inst, S}/(1+z))^2$, where the 
instrumental resolution is $\sigma_{\rm inst, V} = 0.57$ \AA\ for the Valdes et 
al. (2014) templates and $\sigma_{\rm inst, S} = 1.25$ \AA\ for the SDSS 
spectra.
 
Careful consideration should be given to the choice of wavelength range over 
which to perform the fit, with the goal of avoiding regions contaminated by 
strong emission lines and other known sources of systematic bias.  At optical 
wavelengths, the Ca~II $\lambda\lambda$8498, 8542, 8662 infrared triplets are 
optimal for stellar dynamical work because of their relative insensitivity to 
stellar population (Dressler 1984) and AGN contamination (GH06), but they 
are shifted out of the SDSS bandpass and inaccessible to most of our sample.
We do not consider \cahk\ $\lambda\lambda 3934,\ 3968$ because they are blended
with \Ne\ and $\rm H{\epsilon}$ $\lambda$3970 (GH06), the latter being 
especially problematic when a strong A-star population is present.  
After much experimentation, we finally adopt 4100--5400 \AA\ as the fitting 
region, but taking care to avoid strong emission lines (He~II $\lambda 4686$, 
H$\beta$ $\lambda 4861$, \oiii\ $\lambda\lambda 4959, \ 5007$).  We exclude the
\mgb\ $\lambda\lambda5167,\ 5173,\ 5184$ triplets, which suffer from potential 
systematic effects due to [Mg/Fe] enhancement (Barth et al. 2002), but do 
include the G band at 4304 \AA, after masking out H$\gamma$ $\lambda4340$
and \oiii\ $\lambda 4363$.  Figure 2 shows a sample fit.  Our chosen fitting 
regions are similar to those of Greene et al. (2009).  

Kelson et al. (2000) found that \sigs\ is not very sensitive to $P(x)$ and 
$C(x)$, with differences of less than 1\% in \sigs\ for Legendre polynomials 
of different orders.  By contrast, similar tests by GH06 and Barth et al. 
(2002) indicate greater sensitivity of \sigs\ to the choice of polynomial 
order, particularly for fits performed over a relatively narrow wavelength 
range.  Our own tests with the {\tt pPXF} code confirm that the exact choice 
of polynomial order does not have a significant impact on \sigs\ (generally 
\lax\ 2\%--5\%), so long as we avoid very low or high values.  As a compromise,
and for concreteness, we choose a third-order polynomial for $P(x)$ and
$C(x)$.

To verify the robustness of our fitting method, we measured \sigs\ for the
sample of 504 SDSS type 2 AGNs originally used by GH06 (their Figure 2) and 
performed a direct comparison between the two sets of measurements.  The 
agreement is excellent (Figure 3).  Defining the difference between the two 
sets of independent measurements by  $\Delta\sigma_* \equiv 
\log\sigma_*({\rm ours}) - \log\sigma_*({\rm GH06})$, we find a mean value of
$\langle  \Delta \sigma_*  \rangle = 0.002 \pm 0.05$, with little evidence for 
any systematic dependence on \sigs.

\subsubsection{Error Estimates}

The {\tt pPXF} code returns formal (1 $\sigma$) errors of $\sim 13\% \pm 6\%$ 
on \sigs\ for spectra with \sn\  greater than 15 
(Cappellari \& Emsellem 2004).  However, in practice, errors on \sigs\ may be 
dominated by template mismatch or other sources of systematic uncertainties, 
especially for spectra of AGN host galaxies whose continuum may be 
contaminated by other sources of emission.  The situation for type 2 AGNs is 
more favorable than that for type 1 AGNs (e.g., GH06), but even type 2 AGNs 
are not immune.  For instance, the equivalent widths (EWs) of the stellar 
features may be diluted by a featureless continuum from scattered light from 
the nucleus (e.g., Liu et al. 2009; Alexandroff et al. 2013).

We perform a series of Monte Carlo simulations to evaluate the impact of 
template mismatch.  We choose high-\sn\ ($\sim 90$ per pixel near the G band) 
SDSS spectra of six inactive galaxies with \sigs\ $=$ 85, 114, 125, 178, 
220, and  360~\kms, roughly spanning the full range of \sigs\ encountered in 
our sample of type 2 quasars.  In this study, we define the G band over
$4285-4328$~\AA, with the continuum set to the midpoint of a power law
connecting these two end points.  The \sn\ is evaluated in the region
$4400-4600$~\AA, which is relatively free from strong emission or absorption
lines.   We fit each spectrum $10^4$ times, each trial 
randomly varying the set of five template stars (A-type dwarf plus a red giant 
of type F, G, K, and M) drawn from the Indo-U.S. stellar template library of 
Valdes et al. (2004).  Figure~4 shows that the uncertainties on \sigs\ are 
always small, $\sim 4\% \pm 1\%$, with no systematic dependence on \sigs.

The \sn\ of the spectrum has a larger effect on \sigs\ measurements 
(Cappellari \& Emsellem 2004).  The strength (EW) of the stellar 
absorption lines will also matter.  Furthermore, even for spectra of the same 
\sn\ and absorption-line EW, we anticipate that there may be systematic biases 
as a function of \sigs: for the same EW, a larger \sigs\ results in a broader, 
shallower spectral feature, which is more adversely affected by noise 
fluctuations and potential dilution by AGN continuum.  We investigate these
effects as follows.  The above-described set of six high-\sn\ spectra of 
inactive galaxies have EW(G band) $\approx$ 10 \AA.  We dilute the G band from 
the original strength of EW $\approx$ 10 \AA\ to 1 \AA, in steps of 1 \AA, by 
artificially adding a constant continuum level to the spectrum.  Poisson noise 
is introduced to produce a series of spectra with ${\rm S/N} \approx\,3-40$, in 
steps of 1 when ${\rm S/N} < 10$, and in steps of 2 when $10 \leq {\rm S/N} 
\leq 40$.  In total, a grid of 230 pairs of EW and \sn\ was created for each 
of the six original input spectra, and for each pair we simulate 500 
realizations.
 
Figure~5 summarizes the results of these simulations.  As expected intuitively,
the fractional error on \sigs, $\langle \delta \sigma_* \rangle \equiv \langle 
[\sigma_*({\rm input})-\sigma_*({\rm output})]/\sigma_*({\rm input}) \rangle$,
becomes enormous when both \sn\ and EW are low.  On the other hand, 
$\langle \delta \sigma_* \rangle$ \lax\ 15\%$-$20\% when EW \gax\ 6 \AA\ and 
\sn\ \gax\ 10.  The dependence on \sigs\ is weak, especially when \sigs\ \gax\ 
200~\kms.  We utilize the results of Figure~5 to empirically estimate the 
systematic uncertainty for our actual \sigs\ measurements.  The final error 
budget for \sigs\ (Table 1) is the quadrature sum of three components: (1) the 
formal statistical uncertainty from {\tt pPXF}, (2) the systematic uncertainty 
due to potential template mismatch, which, based on the experiments described 
above, we assume to be 4\%, and (3) the statistical uncertainty due to \sn\ 
and EW.

\subsection{Gaseous Velocity Dispersions}

A principal goal of this study is to evaluate whether and how well the velocity 
dispersion of the ionized gas in the narrow-line region of type 2 quasars 
traces the velocity dispersion of the stars in their host galaxies.  Prior to 
measuring the emission lines, the stellar continuum should be properly modeled 
and subtracted, following essentially the same procedure described in Section 3 
(see also Ho et al. 1997a).  The main modification is that the wavelength range 
of the fit now extends from $\sim 3700$ to 7000 \AA\ in order to cover 
\oii\ $\lambda\lambda 3726, 3729$, \oiii\ $\lambda\lambda 4959, 5007$, and 
\sii\ $\lambda\lambda 6716, 6731$, the set of lines we use to measure the 
gaseous velocity dispersion (\sigg).  This template fitting method provides a 
first-order estimate of the global continuum, but it is unlikely to achieve a 
high enough accuracy to yield a robust determination of the local continuum 
for all the emission lines of interest.  We therefore sometimes need to apply 
a second-order correction to adjust the local continuum during the process of 
fitting the individual lines.

The narrow emission lines of AGNs generally have complex shapes that cannot 
be well described by a single Gaussian function.  The profiles of most narrow 
lines, especially higher ionization lines such as \oiii, often have an 
asymmetric blue, and at times red, wing (e.g., Whittle 1985; Veilleux 1991). 
Our sample of type 2 quasars exhibits similar trends.  Although the profile
asymmetry is most obvious for \oiii\ (most likely because of its high \sn), it 
is also frequently discernible for the lower ionization lines of \oii\ and 
\sii\ when the spectra have sufficiently high S/N.  Following standard 
practice (e.g., Ho et al. 1997b; Greene \& Ho 2005a), we fit\footnote{The fits 
are performed using the IDL code {\tt mpfit}.} each line with a single Gaussian
component to represent the ``core'' (\siggc), and, if statistically warranted, 
we add another Gaussian component to reproduce the ``wing'' of the line.  Some 
examples are shown in Figure 6.
The wavelength separations of the three doublets are fixed to their laboratory 
values.  We force the components of each doublet to have the same velocity 
profile, even though in actuality each of the two transitions of the \oii\ and 
\sii\ doublets has slightly different critical densities and hence can have 
mildly different velocity widths (Filippenko \& Halpern 1984; Ho et al. 1996).
We fix the flux ratio of \oiii\ $\lambda$5007/\oiii\ $\lambda$4959 to the 
theoretical value of 3, but for the two densitometers, 
\oii\ $\lambda$3729/\oii\ $\lambda$3726 is allowed to vary from 0.3 to 1.5, 
and \sii\ $\lambda$6716/\sii\ $\lambda$6731 can range between 0.45 and 1.5 
(Osterbrock 1989).  

Apart from the emission-line widths derived from the above-described parametric
fits, we also provide a simpler estimate of line width from a direct, 
non-parametric measurement of the intrinsic FWHM (i.e. after correction for 
instrumental resolution) of the line profile itself.  Then, the 
Gaussian-equivalent velocity dispersion is $\sigma_m \equiv {\rm FWHM}/2.35$.  
Instead of measuring FWHM directly from the data, which is affected by noise, 
it is more effective to calculate it from a model (noise-free) representation 
of the line profile. We construct the model by decomposing the original profile 
using a series of Gaussians, and then summing up the individual components. 
We use as many Gaussians as statistically necessary to fit the profile.  In 
general two Gaussians are sufficient for each of the doublet lines of \oii\ 
and \sii, but \oiii\ may occasionally require up to four to properly account 
for more extended wings.  There is no physical meaning attached to any of 
the Gaussian components; they just serve as a convenient representation of the 
data.  Lastly, as in Greene \& Ho (2005a), we compute the second moment 
(\siggs) of the best-fit model to describe the width of the overall global 
line profile.

The core velocity dispersions and second moments for \oii, \sii, and \oiii, 
corrected for the instrumental resolution of SDSS\footnote{The average 
instrumental spectral resolution of SDSS for our sample is FWHM = $2.54\pm0.30$
\AA\ for \oii, $3.03\pm0.35$ \AA\ for \oiii, and $3.24\pm0.39$ \AA\ for \sii.},
are listed in Table 1.  The error budget for \siggc\ consists of three terms 
combined in quadrature: (1) a formal, statistical uncertainty (generally \lax\ 
5\%) from {\tt mpfit}; (2) a systematic uncertainty due to errors in the 
placement of the local continuum, and (3) a systematic uncertainty due to the 
choice of initial parameters for the multi-component fit.  We estimate terms 
(2) and (3) through a series of Monte Carlo simulations by changing the 
initial fitting parameters.  The error budget for \siggs\ includes only 
terms (2) and (3).

\subsection{External Comparison}

Greene et al. (2009) performed an independent analysis of the Reyes et al. 
(2008) sample, using a methodology very similar that employed in this work.  
They provide \sigs\ measurements for 111 objects, as well as FWHM values for 
\oii\ and \oiii.  Figure 7 directly compares our measurements with those of 
Greene et al. (2009).  The agreement is reasonably good in the mean, although 
the scatter is relatively large for \sigs.  The mean difference between the two
sets of measurements $\langle \Delta\sigma_* \rangle = -0.072\pm0.24$, where
$\Delta\sigma_* \equiv \log\sigma_*({\rm ours})-\log\sigma_*({\rm Greene+09})$.
As for the emission-line widths, $\langle \Delta {\rm FWHM} \rangle =
0.012\pm0.04$ for \oii\ and $-0.002\pm0.03$ for \oiii.

\subsection{Final Set of Measurements}

Out of the original sample of 887 sources, 449 (51\%) have spectra with S/N 
$>$ 3 per pixel and detectable G band with EW larger than 3 \AA. While we 
succeeded to 
measure \sigs\ for most of these, many of the uncertainties are rather large.
For our subsequent analysis, we only make use of the subset of objects with 
\sigs\ measurements that are clearly in excess of the SDSS spectral resolution 
($\sim 70$ \kms) and that have fractional uncertainties $\leq 20\%$. Only 219 
objects (25\% of the original sample) satisfy these criteria.  The number of 
\sigs\ measurements increases to 281 if we relax the fractional uncertainty to 
30\%.  As for the gaseous line widths, after excluding the 126 objects with 
kinematically complex core structure, 594 of the remaining 761 objects have a 
\siggc\ measurement with an intrinsic value larger than 70 \kms\ and a 
fractional uncertainty $\leq 20\%$ for at least one of the three principal 
emission lines (\oii, \oiii, and \sii).  The number of \siggc\ measurements 
increases to 697 if the fractional uncertainty is relaxed to 30\%.  The
majority of the sample has useful measurements of \siggs, at least for \oiii, 
the strongest line, but often also for \oii\ and \sii.  The final set of 
velocity measurements is given in Table 1.  We only list measurements with 
fractional uncertainties less than 30\%.

In total, 142 objects have measurements with fractional uncertainties 
$\leq 20\%$ for both \sigs\ and \siggc\ for at least one emission line.  Not 
surprisingly, they are drawn predominantly from the portion of the parent 
sample with the lowest luminosities (median $\loiii \approx 10^{42.5}\rm 
~erg~s^{-1}$) and redshifts (median $z \approx 0.2$), although on average they 
are still more distant and much more luminous than the sample of type 2 AGNs 
of Greene \& Ho (2005a; Figure 1).

\section{Results}

\subsection{Comparison Between Velocity Dispersions of Stars and Ionized Gas}

Figure 8 begins with a comparison between \sigs\ and \sigg, defined simply as 
FWHM/2.35 (appropriate for a Gaussian function), where FWHM is measured from 
the total line profile of \oii\ (166 objects), \sii\ (154 objects), and \oiii\ 
(169 objects).  It is clear that most of the gas velocities are significantly 
larger than the stellar velocities, and there is little correlation between 
the two quantities.  The average value and standard deviation of 
$\langle (\rm FWHM/2.35)/\sigs \rangle$ is $1.17\pm 0.38, 1.15\pm0.35$, and 
$1.11\pm 0.40$ for \oii, \sii, and \oiii.  This is consistent with the 
conclusion of Greene et al. (2009) and Liu et al. (2009), who parameterized 
the gas velocity dispersion in the same way.  The situation is even more 
extreme for \siggs, the second moment of the line, which is maximally sensitive
to the wings of the profile.  Figure 9a shows that \siggs\ \gax\ \sigs\ for 
nearly all the type 2 quasars, some by very significant amounts.  The range of 
\siggs/\sigs\ spans 0.65 to 6.01, with $\langle \siggs/\sigs \rangle = 1.73 
\pm 0.75$, $1.66 \pm 0.80$, and $1.85\pm 0.88$ for \oii, \sii, and \oiii, 
respectively.  In other words, the narrow lines of type 2 quasars invariably 
have highly non-Gaussian, extended, usually blue asymmetric wings.  Note that 
for lower-luminosity AGNs, only the high-ionization line \oiii\ shows blue 
asymmetry (Greene \& Ho 2005a; Ho 2009; Figure 9a).

By contrast, once we remove the line wings and focus only on the core of the 
emission-line profile (\sigc), the situation improves dramatically (Figure 9b).
The gas velocities become comparable to the stellar velocities, and the scatter
decreases.  We find $\langle \sigc/\sigs \rangle = 1.06\pm0.32$, $0.99\pm0.32$,
and $1.00\pm0.36$ for \oii, \sii, and \oiii.  Because of the small sample 
size, limited dynamic range, and large measurement uncertainties, the 
correlation between \sigs\ and \siggc\ is not statistically strong for the 
type 2 quasars alone.  The Spearman's rank correlation coefficients are $\rho$ 
= 0.32 and 0.34 for \oii\ and \sii, respectively, with a probability for the 
null hypothesis of no correlation of $P_{\rm null} \approx 0.02$; for \oiii, 
$\rho$ = 0.17 and $P_{\rm null} = 0.1$.  The lower luminosity sources of 
Greene \& Ho (2005a)\footnote{As in our analysis, Greene \& Ho (2005a) 
decomposed \oiii\ using two Gaussians, one for the core and another for the 
wing; here we use their core component of \oiii.  They only fit a single 
Gaussian for each of the doublet components of \oii\ and \sii, but in view 
of the relative weakness of the wing component for the low-ionization lines 
in low-luminosity sources (see, e.g., Figure 9a), their single component 
is analogous to the core compoment of type 2 quasars.} behave very similarly: 
$\langle\sigc/\sigs\rangle = 1.13\pm0.38$, $1.11\pm0.35$, and $1.01\pm0.35$. 
For both samples combined, we obtain $\langle\sigc/\sigs\rangle = 1.13\pm0.38$,
$1.11\pm0.35$, and $1.00\pm0.35$.  Now the \sigs$-$\siggc\ relation is 
highly statistically significant, with $\rho = 0.43$, 0.46, and 0.42 and 
$P_{\rm null} < 10^{-6}$ for \oii, \sii, and \oiii.  Among the three emission 
lines considered, \sigc\ for \sii\ formally correlates strongest with \sigs, 
and the ratio of the two has a somewhat smaller scatter.  By comparison, 
\oiii\ exhibits the poorest performance.  As Greene \& Ho (2005a) 
remarked, a possible explanation is that \oiii, which has a significantly 
higher critical density than \oii\ and \sii, originates from closer to the 
nucleus if the narrow-line region is density-stratified and hence has profile 
wings of higher velocity.  

We verified that the above results are quite stable with respect to our 
particular choice of cut on data quality.  Restricting the velocity dispersions
to those with fractional errors $\leq 10\%$ reduces the scatter of 
$\langle\sigc/\sigs\rangle$ from $\sim 0.35$ to $\sim 0.25$ for the type~2 
quasar, at the expense of shrinking the sample, whereas relaxing the fractional 
errors to $30\%$ results in a moderate increase of scatter on 
$\langle\sigc/\sigs\rangle$ but there is little effect on the significance of 
their statistical correlation.  Our analysis shows that, as in lower-luminosity
AGNs (e.g., Nelson \& Whittle 1996; Greene \& Ho 2005a; Ho 2009), the 
kinematics of the narrow-line region gas in type~2 quasars also approximately 
traces the virial motions of the stars in the central regions of their host 
galaxies.  As such, the line widths of the narrow emission lines serve as an 
effective, economical substitute for the bulge stellar velocity dispersion, 
which is difficult, and at times impossible, to measure.  Consistent with 
Greene \& Ho (2005a), we find that \sigg\ only tracks \sigs\ well when care is 
taken to remove the wings of the emission lines.  Bian et al.  (2006) reached 
a similar conclusion for a more limited number of type~2 quasars selected from 
the initial SDSS sample of Zakamska et al. (2003).  For the three emission 
lines investigated here, the low-ionization line \sii\ appears most promising. 
Its core velocity dispersion is, on average, essentially identical to \sigs, 
with a standard deviation of 32\%.  
The next best candidate is \oii, followed by \oiii, as a last resort.  Although
the \oii\ doublet is only marginally resolved by SDSS, and its \siggc\ exhibits
the largest deviation from \sigs, the deviation (6\%) is systematic and 
therefore correctable, and its scatter (32\%) is formally identical to that of 
\sii\ (32\%).  Still, priority should be given to \sii\ whenever possible, as 
\oii\ emanates not only from the narrow-line region but also from extra-nuclear
emission-line regions.  \oii\ is prominent in \hii\ regions (e.g., Gallagher et
al. 1989) and traces galaxy-wide star formation, even in AGN hosts (Ho 2005).

\subsection {Nonvirial Motions: Dependence on AGN Properties}

The vast majority of type 2 quasars emit narrow-line gas with large, 
blueshifted velocities, not only in the high-ionization tracer \oiii, but also
in the low-ionization species \oii\ and \sii.  This is most clearly seen in 
the large, systematic, positive offsets between \siggs\ and \sigs\ (Figure 9a).
While the core component of the emission lines much better traces the 
gravitational potential of the stars (Figure 9b), a small fraction of our 
sample ($\sim 4$\%) still exhibits \oiii\ core velocity dispersions in excess 
of 400 \kms, a couple as high as $\sim 1000$ \kms.  Such velocities certainly 
do not reflect the virial motions of any realistic host galaxy.  The central 
stellar velocity dispersions of nearby galaxies rarely exceed 350 \kms\ (Sheth 
et al. 2003).  Instead, these large, super-virial velocities most likely arise 
from substantial energy injection into the narrow-line region, presumably from 
the AGN, which affect even the core velocities of the narrow emission lines.  
Moreover, while the core gas velocities are close, they are not exactly 
identical, to the stellar velocities.  Do the residuals of the \sigg$-$\sigs\ 
relation depend on AGN properties?  Greene \& Ho (2005a; see also Bian et al. 
2006) find that for \oiii\ the degree of offset between \sigs\ and \siggs, 
which is sensitive to the wing component of the line, statistically scales with
the Eddington ratio.  The wing component is more blueshifted with increasing 
Eddington ratio.  However, even the core of the line profile can be affected.  
Ho (2009) finds that while gravity is the primary agent responsible for the 
velocity broadening of the core of the \nii\ line, the Eddington ratio acts 
as a secondary driver of the \sigc$-$\sigs\ relation: sources with higher
Eddington ratios have systematically larger core line widths relative to \sigs.
Optical AGN luminosity and radio power also correlate with excess gaseous 
velocities, although Eddington ratio appears to be the more dominant factor 
(Ho 2009).  

To evaluate whether these effects also apply to type~2 quasars, we need to 
quantify the BH mass and bolometric luminosity. In the absence of broad 
emission lines, the only viable method to estimate BH masses for type 2 sources
is to make use of empirical scaling relations between BH masses and the 
properties of their host galaxies established for local samples of inactive 
galaxies (Kormendy \& Ho 2013).  The most widely discussed correlations are
those between BH mass and bulge stellar mass and stellar velocity dispersion.  
But challenges arise in connection with our intended current application to the
SDSS-selected sample of type 2 quasars.  First, the BH-bulge scaling relations
(e.g., the \msigma\ relation) has been shown to hold---with reasonable choices
for the virial factor $f$---mainly for nearby ($z \approx 0$) type~1
(broad-line) AGNs (e.g., Onken et al. 2004; Ho \& Kim 2014).  Our sample, by
contrast, consists of type~2 AGNs at moderate redshift.  In order to utilize 
the \msigma\ relation to estimate their BH masses, we must assume (1) that the
\msigma\ relation applies to type~2 AGNs {\it and}\ (2) that we can neglect
(or account for) its possible evolution with redshift, at least for the
redshift range of our sample ($z$ \lax\ 0.8).  Du et al. (2017) investigated 
the virial factors for a handful of Seyfert 2 galaxies with hidden broad-line 
regions and concluded that they are statistically consistent with the average 
virial factors of type 1 AGNs.  This implies that both type 1 and type 2 AGNs 
have broad-line regions of similar structure and kinematics.
In the present context, it suggests---at least tentatively---that the \msigma\
relation can be applied to type 2 AGNs.   The situation regarding the possible
evolution of the \msigma\ relation with redshift is unclear.  On the one hand,
a number of studies suggest that the zero point for the \msigma\
relation increases toward higher redshifts, typically by $\sim 0.3-0.5$ dex
when $z \approx 0.4-0.6$ (Woo et al. 2006, 2008; Treu et al. 2007).  On the
other hand, Shen et al. (2015) argue that this apparent evolution largely
reflects selection effects, while Kormendy \& Ho (2013) show that the latest
calibration of the local \msigma\ relation essentially offsets the previously
reported evolution of zero point at moderate redshifts.  Thus, in the following
analysis, we assume that our sample of low-redshift type 2 quasars obeys the
local \msigma\ relation. From Equation 5 of Kormendy \& Ho (2013),

\begin{equation}
\log \left ( \frac{M_{\rm BH}}{10^9\,M_\odot} \right ) = -(0.500 \pm 0.049) +
(4.429 \pm 0.295) \log \left(\frac{\sigma_{\ast}}{200~{\rm km~s}^{-1}} \right ).
\end{equation}

\noindent
This relation has an intrinsic scatter of 0.28 dex.  Note that the above 
relation applies only to classical bulges and elliptical galaxies, not to 
pseudobulges or to galaxy samples that contain a mixture of bulge 
types\footnote{She et al. (2017) provide a fit for the \msigma\ relation of all 
galaxies with reliable BH masses from Kormendy \& Ho (2013), regardless of 
bulge type: $\log(M_{\rm BH}/10^9\,M_{\sun}) = (-0.68 \pm 0.05) + (5.20\pm0.37)
\log(\sigma/200\; {\rm km\; s}^{-1})$, with an intrinsic scatter of 0.44~dex.}.
Apart from a minority of the objects (Zhao et al. 2018), we do not have 
reliable bulge classifications for most of our sample.  In the subsequent 
analysis, we adopt Equation (2) to derive BH masses, but in Section 4.3 we 
will examine the consequences of relaxing this assumption. 

The \oiii~$\lambda$5007 line offers the most practical means of tracing the
level of AGN activity for the current sample of optically selected type 2
quasars.  It is common practice to use the \oiii\ luminosity to estimate the 
AGN bolometric luminosity, but considerable debate surrounds the proper 
bolometric correction, the role of extinction correction, and whether or not 
the bolometric correction is luminosity-dependent (e.g., Heckman et al. 2004; 
Netzer et al. 2006; Kauffmann \& Heckman 2009; Lamastra et al. 2009; Stern \& 
Laor 2012; Pennell et al. 2017; see discussion in Heckman \& Best 2014). For 
concreteness, we adopt a constant bolometric correction of 600 (i.e., 
$L_{\rm bol} = 600 \loiii$) to the extinction-corrected \oiii\ luminosity, 
following the recommendation of Kauffmann \& Heckman (2009), with an 
uncertainty of $\pm 150$.  For comparison, we will also consider the constant 
bolometric correction of 3500 for extinction-uncorrected \oiii\ luminosity 
originally advocated by Heckman et al. (2004), as well as the 
luminosity-dependent, extinction-corrected bolometric corrections of Lamastra 
et al. (2009), as parameterized by Trump et al. (2015).  We estimate the 
internal extinction from the observed Balmer decrement, assuming an intrinsic 
value of \halpha/\hbeta\ = 3.1 appropriate for AGNs (Halpern \& Steiner 1983) 
and the extinction curve of Cardelli et al. (1989).  For objects for which the 
Balmer decrement could not be measured, we adopt the median value of 
\halpha/\hbeta\ = 4.04 obtained from the objects for which this quantity was 
measured.

Figure 10a shows the relation between $\Delta\sigma^s \equiv \log\sigma^s_{g} - 
\log\sigma_{\ast}$ and \lledd, both for our sample of type~2 quasars (red 
points) and for lower luminosity sample of Seyferts from Greene \& Ho (2005a; 
small grey points).  We consistently recalculate the black hole masses and 
bolometric luminosities of the Greene \& Ho objects using the \msigma\ relation
adopted in this work (Equation 2) and the bolometric correction of 600 for 
extinction-corrected \oiii\ luminosity\footnote{The \oiii\ luminosities of 
Greene \& Ho (2005a) were not corrected for internal extinction. For consistency
with the conventions of our study, we applied an internal extinction 
correction to the luminosities of Greene \& Ho (2005a) using Balmer decrements 
for their objects obtained from the JHU/MPA database ({\url https://wwwmpa.mpa-garching.mpg.de/SDSS)}}.  The velocity residuals of all three transitions 
correlate positively and strongly with the Eddington ratio, both for the
high-luminosity and low-luminosity samples.  The combination
of the lower luminosity Greene \& Ho sources and our high-luminosity quasars 
provides the largest dynamic range in Eddington ratio, yielding the following 
functional dependence between $\Delta\sigma^s$ and $L_{\rm bol}/L_{\rm Edd}$ 
(black solid line in Figure 10a): 

\begin{equation}
\Delta\sigma^s\! =\!
\left\{\!
\begin{array}{ll}
\!(0.106\pm0.004)~\log (L_{\rm bol}/L_{\rm Edd}) \, + \, (0.255\pm0.009) & \!\rm{for~[O~{II}]} \\
\!(0.093\pm0.004)~\log (L_{\rm bol}/L_{\rm Edd}) \, + \, (0.222\pm0.009) & \!\rm{for~[S~{II}]} \\
\!(0.149\pm0.005)~\log (L_{\rm bol}/L_{\rm Edd}) \, + \, (0.392\pm0.010) & \!\rm{for~[O~{III}]},
\end{array}
\right.
\end{equation}
 
\noindent
with Spearman's rank correlation coefficients of $\rho$ = 0.43, 0.38, and 0.56, 
respectively, and $P_{\rm null} < 10^{-6}$.   Greene \& Ho (2005a) 
already presented the results for \oiii\ for type 2 Seyferts.  Here we confirm 
that it extends to sources of even higher luminosity and Eddington ratio, and, 
moreover, that it also applies to the low-ionization lines \oii\ and \sii. 

More surprisingly, we find that even the {\it core}\ component of the lines 
are affected systematicaly by Eddington ratio (Figure 10b).  The type~2 
quasars exhibit the strongest correlation between $\Delta\sigma^c$ and \lledd\ 
($\rho$ = 0.66, 0.49, and 0.61 for \oii, \sii, and \oiii; $P_{\rm null} < 
10^{-6}$), but statistically significant correlations also exist for the 
lower-luminosity objects of Greene \& Ho (2005a) alone ($\rho$ = 0.36, 0.32, 
and 0.14 for \oii, \sii, and \oiii; $P_{\rm null} < 10^{-6}$), and for both 
samples combined ($\rho$ = 0.33, 0.26, and 0.13 for \oii, \sii, and \oiii; 
$P_{\rm null} < 10^{-6}$).  The sensitivity of \siggc\ to \lledd\ echoes the 
findings of Ho (2009; overplotted as dotted line in the \oii\ and \sii\ panels 
of Figure 10b) based on the low-ionization line \nii\ \lamb6583, which extend 
the Eddington ratios down by $\sim 2$ orders of magnitude, to \lledd\ $\approx\,
10^{-5.5}$.  

We looked for, but failed to find, any dependence between velocity excess 
(either $\Delta\sigma^s$ or $\Delta\sigma^c$) and \oiii\ luminosity 
or 20~cm radio power (available from the FIRST database; Becker et al. 1995).  
Greene \& Ho (2005a) arrived at a similar conclusion for $\Delta\sigma^s$.  
The mildly significant correlation between \delsigc(\nii) and H$\alpha$ 
luminosity and radio power seen by Ho (2009), which he considers to be 
secondary to the influence of Eddington ratio, may be masked by the 
larger measurement uncertainties of the present data.

\subsection {Black Hole Masses and Eddington Ratios}

One of the main objectives of this study is to evaluate the BH masses and 
Eddington ratios of type 2 quasars. We estimate BH masses using the \msigma\ 
relation (Equation 2).  Having established that the kinematics of the 
narrow-line region in type 2 quasars largely tracks the virial velocities of 
the bulge, we use the velocity dispersion of the core component (\sigc) of the 
narrow lines to estimate \sigs\ whenever the latter is unavailable.  We give 
preference to \sigc\ from \sii, followed by \oii, and then \oiii.  We adjust 
the gas velocity dispersions by applying a small statistical zeropoint scaling 
correction: $\langle \sigc/\sigs \rangle = 0.99$, 1.06, and 1.00 for \sii, 
\oii, and \oiii.  As mentioned in Section 4.2, a small fraction of the objects 
have exceptionally large \oiii\ line widths of clearly non-gravitational 
origin.  We omit these from the following analysis.  

The BH masses of type~2 quasars range from $M_{\rm BH} \approx 10^{6.5}$ to 
$10^{10.4} \, M_\odot$, with a median value of $10^{8.2} \, M_\odot$ 
(Figure~11a).  Had we adopted the \msigma\ relation of She et al. (2017; see 
footnote 7), which does not distinguish between classical and pseudo bulges, 
the median BH mass would decrease by $\sim 0.25$ dex. 
%
%
The Eddington ratios span $L_{\rm bol}/L_{\rm Edd} \approx 10^{-2.9}$ to 
$10^{1.8}$, with a median value of $10^{-0.7}$ (Figure~11b).  These BH masses
and Eddington ratios are very similar to the results of Greene et al. (2009).
The alternative \oiii\ bolometric corrections of Trump et al. (2015) yield a 
lower median $L_{\rm bol}/L_{\rm Edd}$ lower by $\sim 0.2$ dex, while Heckman 
et al.'s (2004) bolometric correction would lead to an increase of $\sim 0.4$ 
dex in the median $L_{\rm bol}/L_{\rm Edd}$.  Figure 12 compares type~2 quasars 
with optically selected type~1 AGNs over the same redshift range ($z < 0.83$) 
studied by Shen et al. (2011). Roughly half of the type~1 sources have $\loiii 
\ga 2 \times 10^{42}$ \lum, the lower limit threshold adopted for type~2 
quasars (Reyes et al. 2008).  The BH masses for the type~1 sources, taken 
directly from Shen et al. (2011), are based on single-epoch virial mass 
estimators using the broad H$\beta$ or Mg~II emission lines.  For consistency 
with the type~2 sources, we also compute the bolometric luminosities of the 
type~1 sample using \loiii\ from Shen et al., but in this instance we need to 
use the \oiii\ bolometric correction of Heckman et al. (2004) because 
extinction corrections are not available.  Taken at face value, type 2 quasars 
cover a similar range of BH masses and Eddington ratios as type 1 quasars, with
type 1 systems tending toward somewhat higher $M_{\rm BH}$ (median $10^{8.5}\, 
M_\odot$) and lower $L_{\rm bol}/L_{\rm Edd}$ (median $10^{-0.9}$).
%
%
%
%
%
%
The lower-luminosity type 2 Seyferts of Greene \& Ho (2005a) differ more
dramatically from type 2 quasars (median $M_{\rm BH} = 10^{7.9} \, M_\odot$; 
median $L_{\rm bol}/L_{\rm Edd} = 10^{-2.1}$), but this is a trivial 
consequence of sample selection (the optically selected type 2 quasars are
required to meet the luminosity threshold of quasars).

Significant uncertainties and selection effects complicate the comparison 
between type 1 and type 2 quasars (e.g., BH mass estimation, bolometric 
corrections, dust extinction, etc), and  we do not recommend taking too 
literally the numerical values for any individual object or of the apparent 
minor differences between the two quasar populations.  Nevertheless, it is 
worth remarking that the bolometric luminosities for $\sim 20\%$ of the 
type~2 quasars formally exceed the Eddington limit, some by a significant 
margin; 11 objects (1.6\%) have $L_{\rm bol}/L_{\rm Edd} \geq\ 10$.  The 
sources with the most extreme Eddington ratios have the highest luminosities 
(\loiii\ \gax\ $10^{43.5}$ \lum\ and lowest BH masses ($M_{\rm BH} \approx 
10^{6.5}-10^{7.5}$ \solmass).  Type~2 quasars, as a population, appear to have 
somewhat higher accretion rates than their type~1 counterparts, and the subset 
with the least massive BHs are radiating at highly super-Eddington rates.  
This may arise naturally in the evolutionary scenario linking obscured and 
unobscured AGNs.

\section{Summary}

We analyze the optical spectra of the 887 low-redshift ($0.037 < z < 0.83$) 
type~2 quasars from the SDSS catalog of Reyes et al. (2008), with the aim of 
deriving two fundamental physical parameters for this population, their BH 
mass and Eddington ratio.  Our strategy is to estimate BH masses using the 
\msigma\ relation, employing directly measured central stellar velocity 
dispersions whenever possible and otherwise indirectly via the velocity 
dispersions of the bright narrow lines \oii, \sii, and \oiii.

Our main results are as follows:

\begin{enumerate}

\item The kinematics of the core component of the narrow emission lines trace 
the gravitational potential of the stars of the host galaxy, albeit with 
significant scatter.  The low-ionization lines \sii\ and \oii\ are most 
effective, but even the high-ionization line \oiii\ can be used.

\item The AGN plays a secondary but still important role in governing the 
kinematics of narrow-line region.  Narrow emission lines of both low and high 
ionization exhibit significantly blueshifted velocity wings.  The magnitude of 
these nonvirial motions becomes more prominent with increasing Eddington ratio.
  
\item Optically selected type 2 quasars have BH masses spanning $M_{\rm BH} 
\approx 10^{6.5}$ to $10^{10.4} \, M_\odot$ (median $10^{8.2} \, M_\odot$) and 
Eddington ratios \lledd\ $\approx$ $10^{-2.9}$ to $10^{1.8}$ (median 
$10^{-0.7}$).  A minosity exceeds the Eddington limit by more than a factor 
of 10.
 
\item The accretion rates of type 2 quasars are somewhat higher than those of 
type 1 quasars of similar redshift.  
\end{enumerate}

\acknowledgements
This work was supported by the National Key R\&D Program of China 
(2016YFA0400702), the National Science Foundation of China (11473002, 11721303),
the National Youth Fund (11303008), and the Youth Foundation of Hebei Province 
of China (A2011205067).  We thank Jenny Greene for sending the data from 
Greene \& Ho (2005a). MK is grateful to Jian-Min Wang for his hospitality 
and help during her visit to IHEP and to JinLin Han for his help during her 
stay at NAOC. She thanks Song Huang, Ligang Hou, Xianmin Meng, Jinyi Shangguan, Jiayi Sun, and Junzhi Wang for their help.

Funding for the SDSS has been provided by the Alfred P. Sloan Foundation, the Participating Institutions, the National Science Foundation, the U.S. Department of Energy, the National Aeronautics and Space Administration, the Japanese Monbukagakusho, the Max Planck Society, and the Higher Education Funding Council for England. The SDSS Web Site is {\url http://www.sdss.org/}

The SDSS is managed by the Astrophysical Research Consortium for the Participating Institutions. The Participating Institutions are the American Museum of Natural History, Astrophysical Institute Potsdam, University of Basel, University of Cambridge, Case Western Reserve University, University of Chicago, Drexel University, Fermilab, the Institute for Advanced Study, the Japan Participation Group, Johns Hopkins University, the Joint Institute for Nuclear Astrophysics, the Kavli Institute for Particle Astrophysics and Cosmology, the Korean Scientist Group, the Chinese Academy of Sciences (LAMOST), Los Alamos National Laboratory, the Max-Planck-Institute for Astronomy (MPIA), the Max-Planck-Institute for Astrophysics (MPA), New Mexico State University, Ohio State University, University of Pittsburgh, University of Portsmouth, Princeton University, the United States Naval Observatory, and the University of Washington.


\begin{deluxetable}{lccccccccccccccccc}
\setlength{\tabcolsep}{1.5pt}
\tablecaption{Sample Properties} 
\tablecolumns{16}
\tablewidth{490pt}
\tabletypesize{\tiny}
\tablehead{
\colhead{Object Name} &
\colhead{$z$} & 
\colhead{$A_V$} & \colhead{S/N}&
\colhead{EW}&
\colhead{H$\alpha$/H$\beta$}&
\colhead{$L_{\rm [O~{III]}}$}&
\colhead{$\sigma_{\ast}$} &
\colhead{$\sigma^c_{\rm [S~ {II]}}$}&
\colhead{$\sigma^c_{\rm [O~ {II]}}$}&
\colhead{$\sigma^c_{\rm [O~ {III]}}$}&
\colhead{$\sigma^s_{\rm [S~ {II]}}$}&
\colhead{$\sigma^s_{\rm [O~ {II]}}$}&
\colhead{$\sigma^s_{\rm [O~ {III]}}$}&
\colhead{$M_{\rm BH}$}&
\colhead{$L_{\rm bol}/L_{\rm Edd}$}\\
\colhead{SDSS}& 
\colhead{}& 
\colhead{(mag)}& 
\colhead{}& 
\colhead{(\AA)}&
\colhead{}& 
\colhead{($\rm erg~s^{-1}$)}&
\colhead{(\kms)}&
\colhead{(\kms)}&
\colhead{(\kms)}&
\colhead{(\kms)}&
\colhead{(\kms)}&
\colhead{(\kms)}&
\colhead{(\kms)}&
\colhead{($M_{\sun}$)}& 
\colhead{}\\
 \colhead{(1)}& \colhead{(2)} & \colhead{(3)}&\colhead{(4)}&\colhead{(5)}&\colhead{(6)}&\colhead{(7)}& \colhead{(8)} & \colhead{(9)} & \colhead{(10)} & \colhead{(11)}& \colhead{(12)} & \colhead{(13)}& \colhead{(14)}& \colhead{(15)} & \colhead{(16)}}
\startdata
J000259.10$+004018.1$       &  0.6007  &  0.0762  &  4.1   &  < 3   &    4.04 &   43.13$\pm$0.05    &  \nd                &    \nd              &    \nd              &    136.0$\pm$23.2 &         \nd         &    343.1$\pm$5.3    &    464.9$\pm$28.6   &    7.76$\pm$0.80    &    0.05$\pm$0.81    \\
J000351.83$-010142.0$       &  0.2689  &  0.0968  &  7.1   &  5.4   &    4.04 &   42.19$\pm$0.05    &    241.7$\pm$65.5   &    \nd              &    \nd              &    225.4$\pm$63.7 &         \nd         &    284.9$\pm$27.7   &    276.9$\pm$32.4   &    \nd              &    \nd              \\
J000453.43$-005038.4$       &  0.6430  &  0.1100  &  3.1   &  < 3   &    4.04 &   43.31$\pm$0.05    &  \nd                &    \nd              &    137.3$\pm$9.8    &    174.0$\pm$2.0 &         \nd         &    382.6$\pm$15.8   &    500.5$\pm$42.0   &    7.66$\pm$0.75    &    0.33$\pm$0.76    \\
J000728.83$+010604.0^{a}$   &  0.4663  &  0.0988  &  7.8   &  6.0   &    4.04 &   42.37$\pm$0.05    &    203.4$\pm$50.6   &    \nd              &    \nd              &    \nd &         \nd         &    599.3$\pm$104.7  &    402.0$\pm$93.1   &    \nd              &    \nd              \\
J001000.75$-011008.5$       &  0.5823  &  0.0936  &  6.1   &  < 3   &    4.04 &   42.71$\pm$0.06    &  \nd                &    \nd              &    392.6$\pm$57.1   &    204.4$\pm$59.0 &         \nd         &    504.4$\pm$41.9   &    260.1$\pm$13.8   &    9.69$\pm$0.79    &    $-$2.30$\pm$0.80 \\
J001111.95$+005626.3$       &  0.4094  &  0.0812  &  6.9   &  < 3   &    4.04 &   42.69$\pm$0.05    &  \nd                &    \nd              &    133.6$\pm$17.5   &    138.7$\pm$31.6 &         \nd         &    414.3$\pm$43.0   &    289.6$\pm$19.4   &    7.61$\pm$0.78    &    $-$0.25$\pm$0.79 \\
J001126.95$+155329.6$       &  0.0999  &  0.2062  &  12.3  &  5.5   &    3.76$\pm$0.05 &   42.14$\pm$0.02    &    154.9$\pm$19.4   &    141.2$\pm$5.9    &    159.1$\pm$8.2    &    134.9$\pm$5.5 &    247.6$\pm$31.3   &    225.5$\pm$18.2   &    224.0$\pm$6.7    &    8.01$\pm$0.37    &    $-$1.19$\pm$0.39 \\
J001206.31$-094725.6$       &  0.1668  &  0.1057  &  13.8  &  5.2   &    5.17$\pm$0.22 &   42.92$\pm$0.06    &    146.7$\pm$21.2   &    114.0$\pm$7.4    &    167.5$\pm$43.9   &    154.0$\pm$7.4 &    242.0$\pm$11.5   &    423.1$\pm$34.3   &    285.0$\pm$18.4   &    7.90$\pm$0.40    &    $-$0.30$\pm$0.42 \\
J002016.88$-093244.7$       &  0.3600  &  0.1289  &  20.9  &  < 3   &    5.17$\pm$0.31 &   43.06$\pm$0.08    &  \nd                &    \nd              &    182.9$\pm$8.5    &    176.0$\pm$4.2 &         \nd         &    376.3$\pm$31.7   &    469.9$\pm$4.2    &    8.22$\pm$0.74    &    $-$0.48$\pm$0.75 \\
\enddata
\tablecomments{\\
$^{a}$Kinematically complex emission-line core.\\
Column (1): SDSS name.
Column (2): Redshift.
Column (3): Extinction in the $V$ band.
Column (4): Average S/N per pixel over the region $4400-4600$~\AA. 
Column (5): EW of the G band.
Column (6): Observed Balmer decrement used to estimate internal extinction. 
Objects for which the Balmer decrement could not be reliably determined are 
assigned a value of H$\alpha$/H$\beta$ = 4.04, the median of the detected 
objects.
Column (7): Extinction-corrected luminosity of [O {III}] $\lambda$5007. 
The uncertainties include measurement error and uncertainties of the Balmer 
decrement.
Column (8): Stellar velocity dispersion.
Columns (9)--(11): Gaseous velocity dispersion of the core component of 
[S \tiny {II}] $\lambda\lambda$6716, 6731,  
[O \tiny {II}] $\lambda\lambda$3726, 3729, and 
[O \tiny {III}] $\lambda$5007.
Columns (12)--(14): Second moment of the line profile of
[S \tiny {II}] $\lambda\lambda$6716, 6731, 
[O \tiny {II}] $\lambda\lambda$3726, 3729, and
[O \tiny {III}] $\lambda$5007.
Column (15): BH mass estimated from the $M_{\rm BH}-\sigma_\ast$ relation of 
Kormendy \& Ho (2013; Equation 2), using, in the order of preference, 
$\sigma_\ast$, $\sigma^c_{\rm [S~ {II}]}$, 
$\sigma^c_{\rm [O~ {II}]}$, and $\sigma^c_{\rm [O~ {III}]}$.
with \siggc\ statistically corrected as described in Section 4.3.
Column (16): Eddington ratio calculated with extinction-corrected 
[O  {III}] luminosity (Col. 7) and BH mass (Col. 15). Uncertainties 
introduced by the bolometric correction are included in the error budget.
{\it This table is available in its entirety in a machine-readable form in the 
on-line journal.  A portion is shown here for guidance regarding its form and 
content.}
}
\end{deluxetable}

\clearpage

\vskip 3.1cm
\hskip -1.0cm
\begin{figure}
\plotone{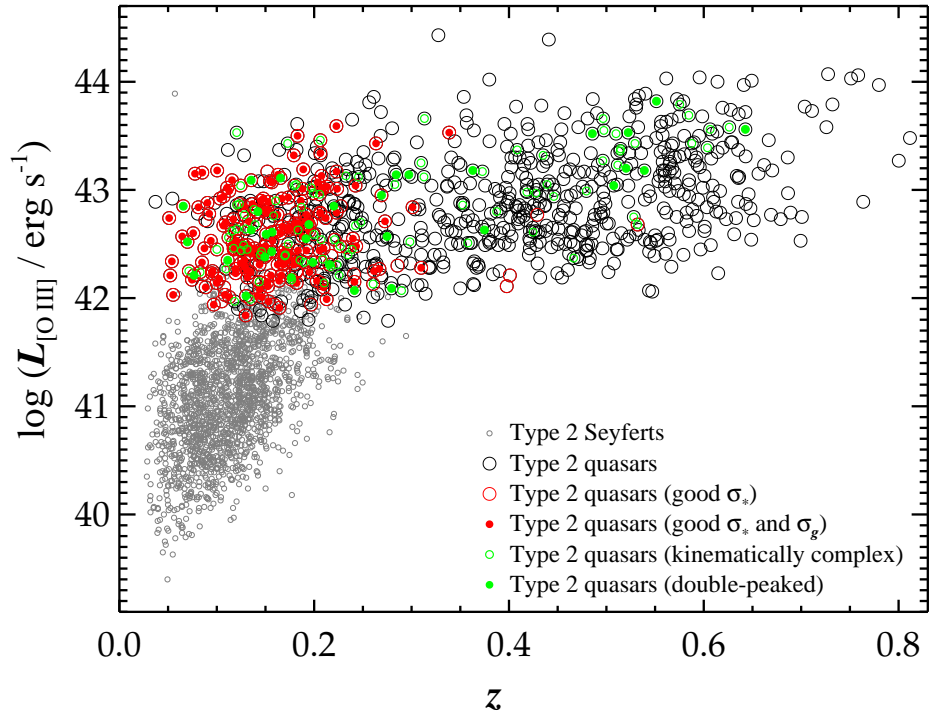}

\caption{
Distribution of \oiii\ luminosities and redshifts of our parent sample of 887 
type~2 quasars (open black symbols).  Plotted in red are the 219 objects that 
have $\sigma_{\ast}$ measurements with fractional errors $\leq 20$\%, with the 
final sample of 142 objects having robust values of both $\sigma_{\ast}$ and 
$\sigma_g$ further highlighted as filled red symbols.  Green symbols denote 
objects with kinematically complex emission-line cores, and green filled 
symbols mark the subset with double-peaked profiles.  The small grey points 
are the lower luminosity type~2 AGNs (Seyferts) from Greene \& Ho (2005a).}
\end{figure}

\vskip 0.1cm

\hskip -0.5cm
\vskip 1.1cm
\hskip -0.1cm
\begin{figure}
\includegraphics[width=14cm, height=10cm,angle=0]{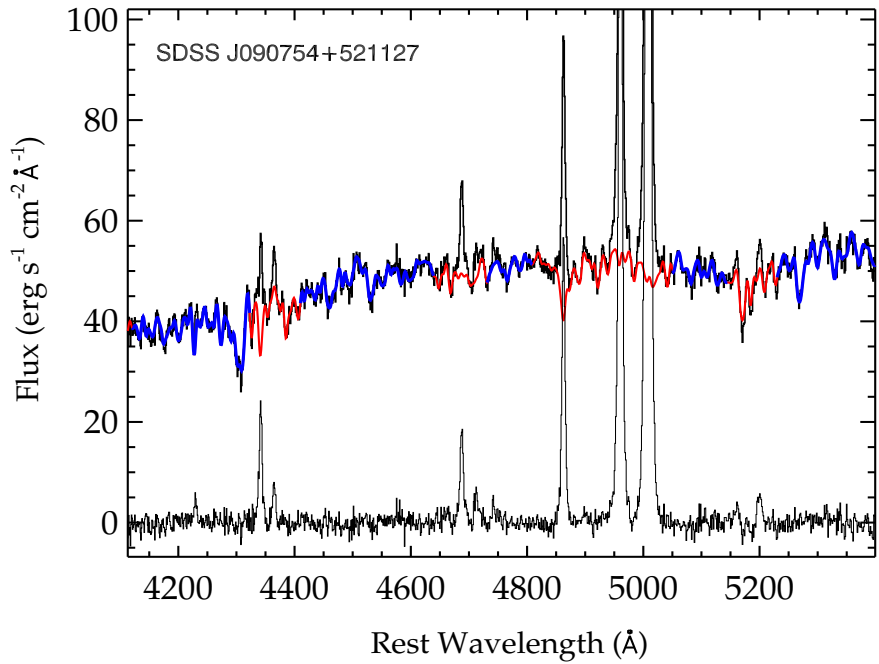}
\figcaption[fig2.ps]{
Illustration of the method for \sigs\ measurement.  The SDSS spectrum, 
corrected for extinction and redshift, is plotted as thick black histograms.
The best-fit model, a scaled, broadened composite stellar template comprising 
several individual stars (F, K, and M late-type red giants), is plotted in 
blue.  Regions of the spectrum containing strong emissions are excluded from 
the fit (red).  The residuals (data minus best-fit model) are plotted on 
the bottom. 
\label{fig2}}
\end{figure}

\vskip .6cm
\hskip 2.0cm
\begin{figure}
\begin{center}
\includegraphics[width=9cm, height=9cm,angle=0]{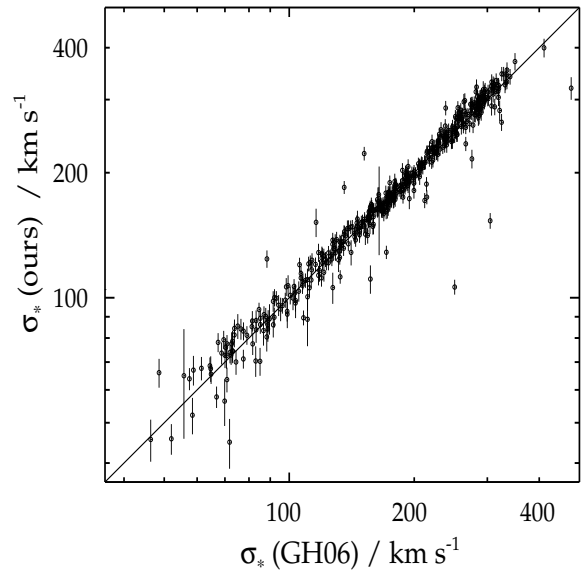}

\figcaption[fig3.ps]{
Comparison between our measurements of \sigs\ and those of GH06 for their 
reference sample of 504 active galaxies.  The agreement is excellent.  The mean 
difference between the two sets of measurements is $\langle \Delta\sigma_* 
\rangle = 0.002\pm0.05$, where $\Delta\sigma_* \equiv \log\sigma_*({\rm ours}) 
- \log\sigma_*({\rm GH06})$.
\label{fig3}}
\end{center}
\end{figure}

\vskip 0.2cm
\hskip -0.5cm
\begin{figure}
\includegraphics[width=16cm, height=12cm,angle=0]{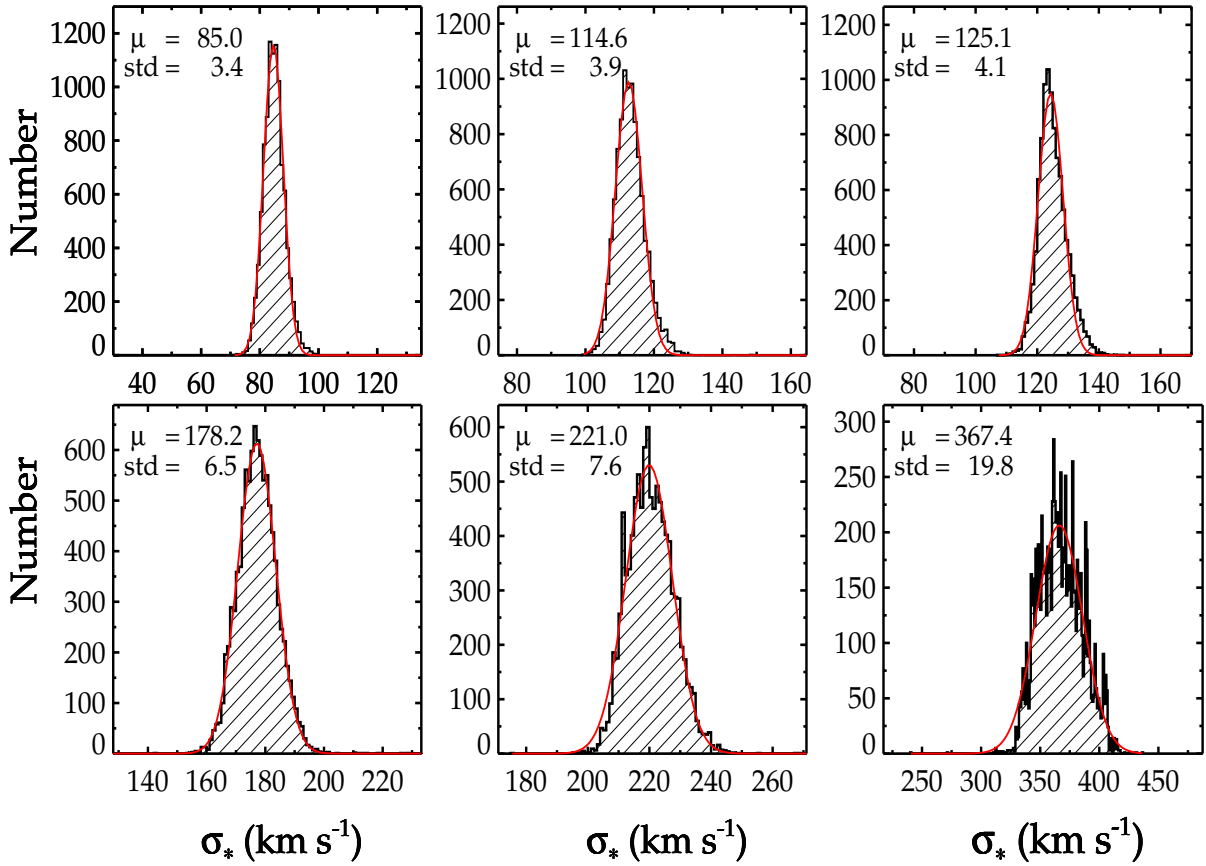}

\figcaption[fig4.ps]{
The effect of template mismatch on measured velocity dispersions. For each of 
six high-S/N (\gax 90) SDSS galaxy spectra with \sigs\ $=$ 85, 114, 125, 
178, 220, and 360 \kms, we perform $10^4$ fits with different stellar templates
constructed from random combinations of stars with spectral type A, F, G, K, 
and M. The mean ($\mu$) of the resulting distribution of \sigs\ and its 
standard deviation from a Gaussian fit (red line) to the distribution are 
given in the top left corner of each panel.
\label{fig4}}
\end{figure}

\vskip 0.6cm
\hskip -0.5cm
\begin{figure}
\includegraphics[width=16cm, height=12cm,angle=0]{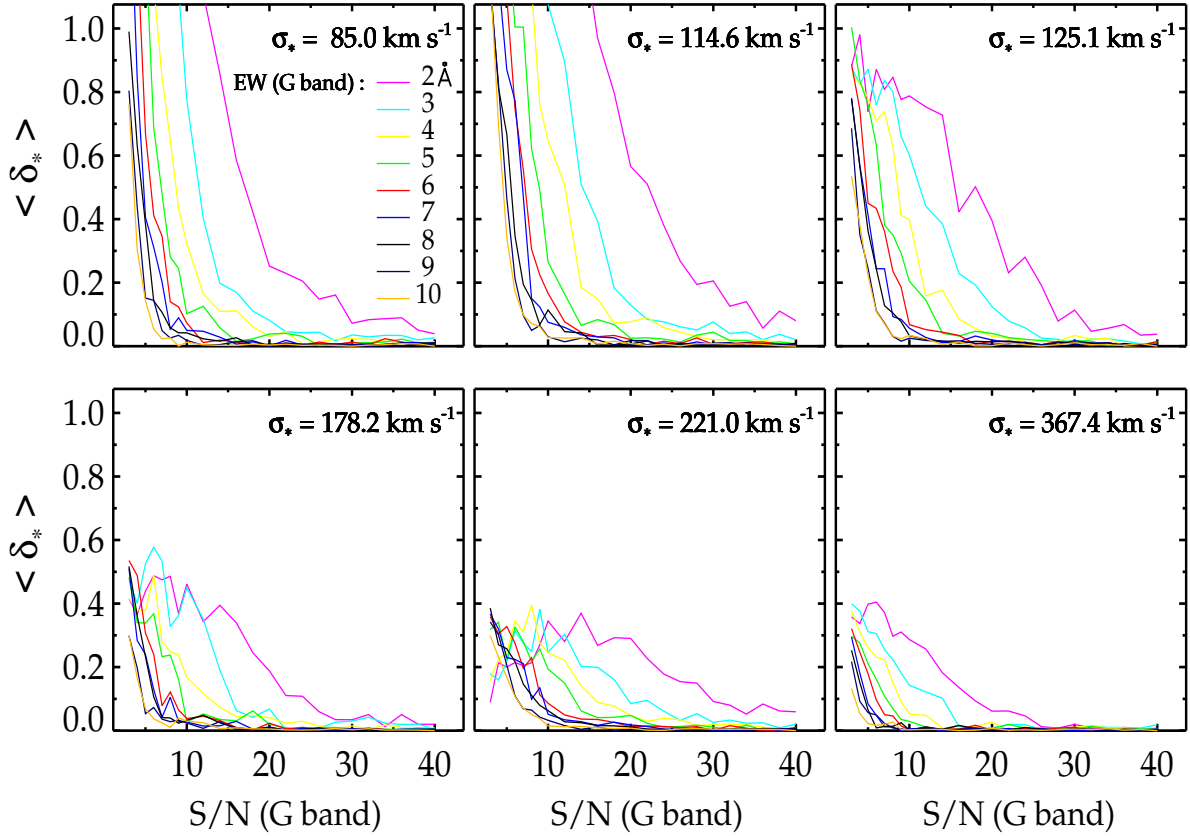}

\figcaption[fig5.ps]{
Systematic errors on the measured stellar velocity dispersions introduced by 
S/N and strength (EW) of the G band stellar feature.  The panels show 
simulations performed on six high-S/N (\gax 90) SDSS galaxy spectra with 
\sigs\ $=$ 85, 114, 125, 178, 220, and 360 \kms.  The mean fractional error on 
\sigs\ is given as $\langle \delta \sigma_* \rangle = \langle 
[\sigma_*({\rm input})-\sigma_*({\rm output})]/\sigma_*({\rm input}) \rangle$,
for different combinations of S/N and EW(G band).
\label{fig5}}
\end{figure}

\vskip 3.1cm
\hskip -1.0cm
\begin{figure}
\includegraphics[width=16cm, height=12cm,angle=0]{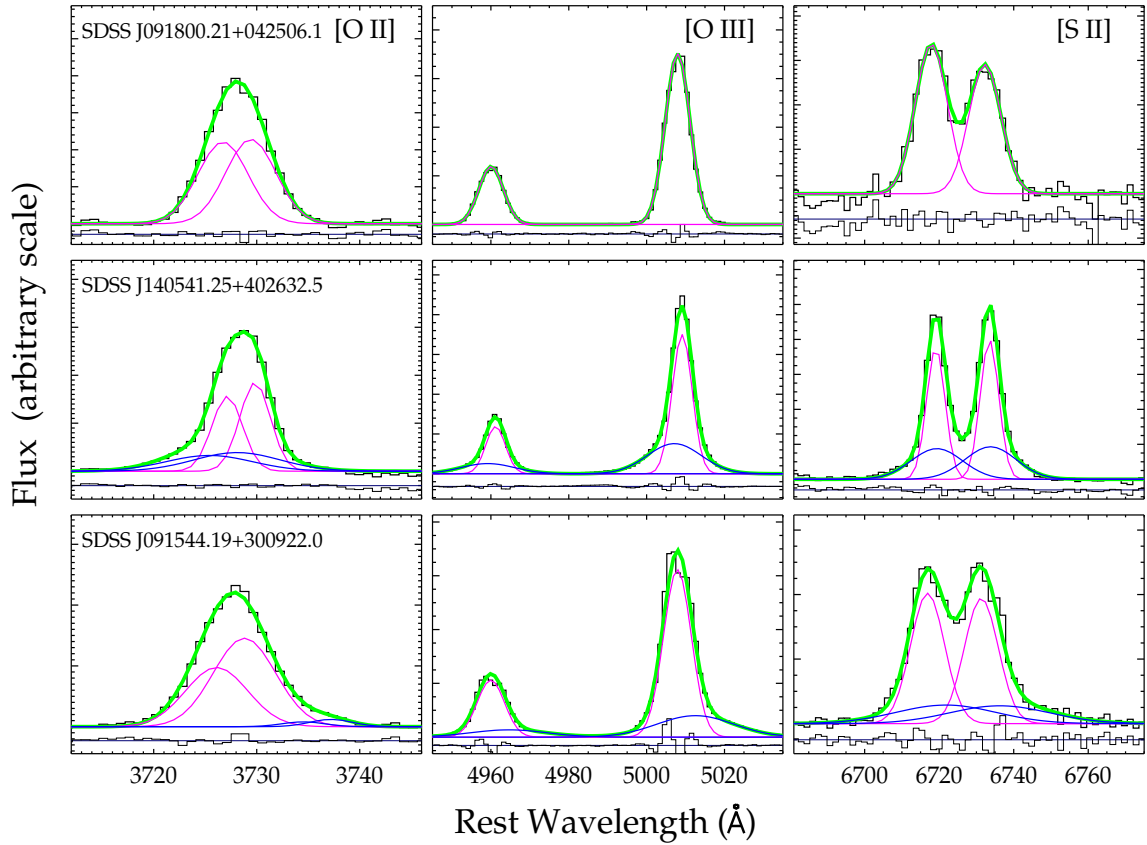}

\figcaption[fig6.ps]{
Sample fits for the emission lines (left) \oii\ $\lambda\lambda 3726, 
3729$, (middle) \oiii\ $\lambda\lambda 4959, 5007$, and (right) 
\sii\ $\lambda\lambda 6716, 6731$.  The top object requires only a single 
Gaussian component for each line, whereas the lines for the other two objects 
each require two components, one whose wing is blueshifted (middle row) and 
the other that is redshifted (bottom row).  The data are plotted in black, 
the best-fit model in green, and individual components in magenta (core) and 
blue (wing).  The residuals (data minus model) are plotted on the bottom 
of each panel as thin black histograms.
\label{fig6}}
\end{figure}

\begin{figure}
\includegraphics[width=16cm, angle=0]{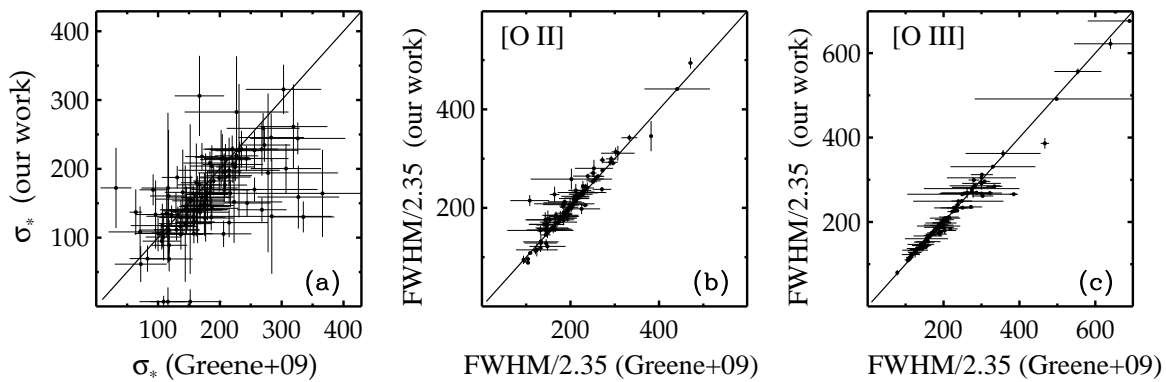}

\figcaption[fig7.ps]{
Comparison of our measurements with those of Greene et al. (2009) for (a) 
\sigs, (b) FWHM of \oii, and (c) FWHM of \oiii.  The solid line denotes 
the 1:1 relation.
\label{fig7}}
\vskip 0.0cm
\end{figure}

\begin{figure}
\includegraphics[width=14cm, angle=0]{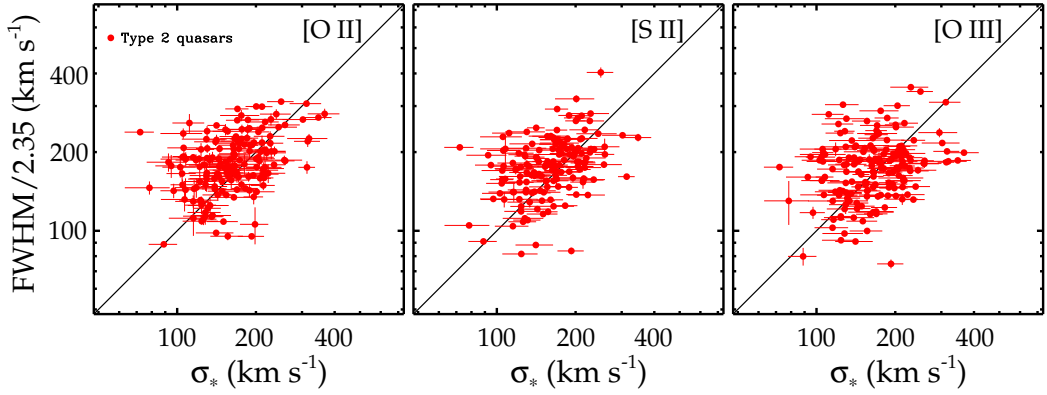}

\vskip -6.1cm
\figcaption[fig8.ps]{
Comparison of stellar velocity dispersions \sigs\ with gaseous velocity
dispersions estimated from FWHM/2.35 of the \oii, \sii, and \oiii\ narrow
emission lines.  
The data consist of the subset of type~2 quasars in this study having 
velocity dispersions measurements with fractional uncertainties $\leq 20\%$.  
The solid line denotes the 1:1 relation.
\label{fig8}}
\vskip 0.0cm
\end{figure}

\hskip -0.5cm
\begin{figure}
\epsscale{0.9}
\plotone{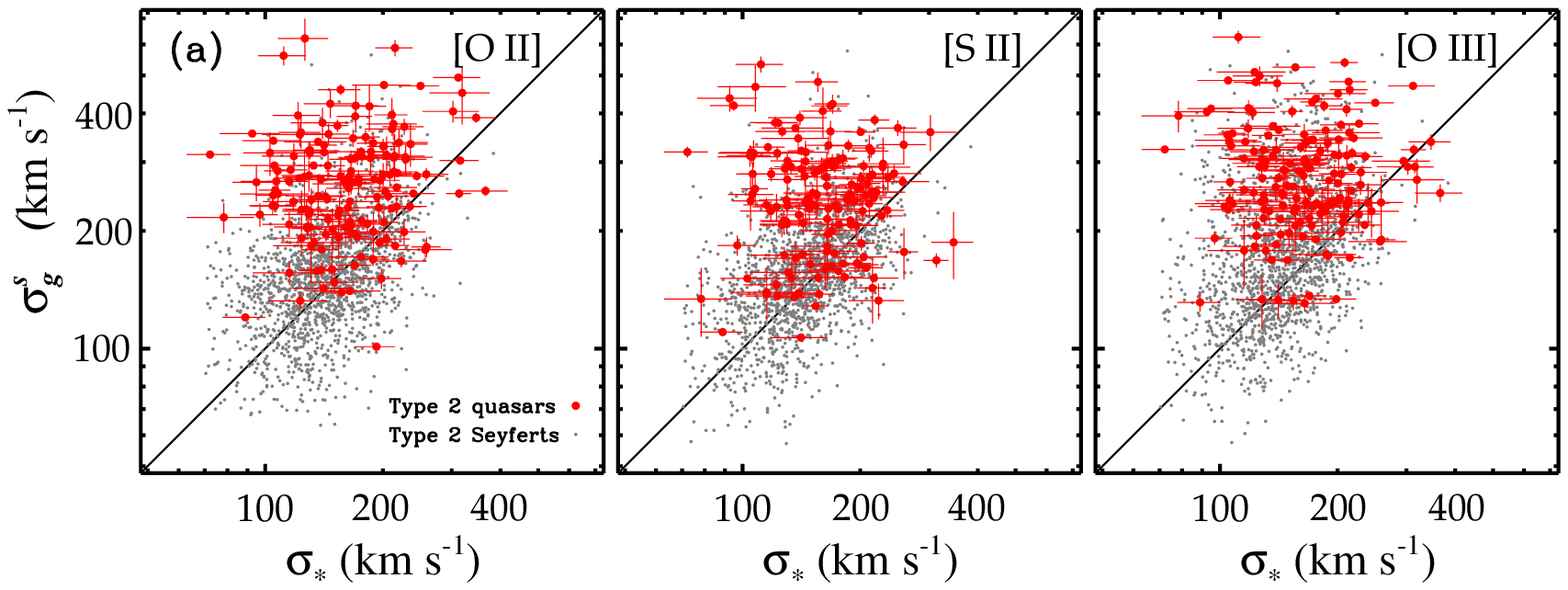}
\vskip -7.0cm
\plotone{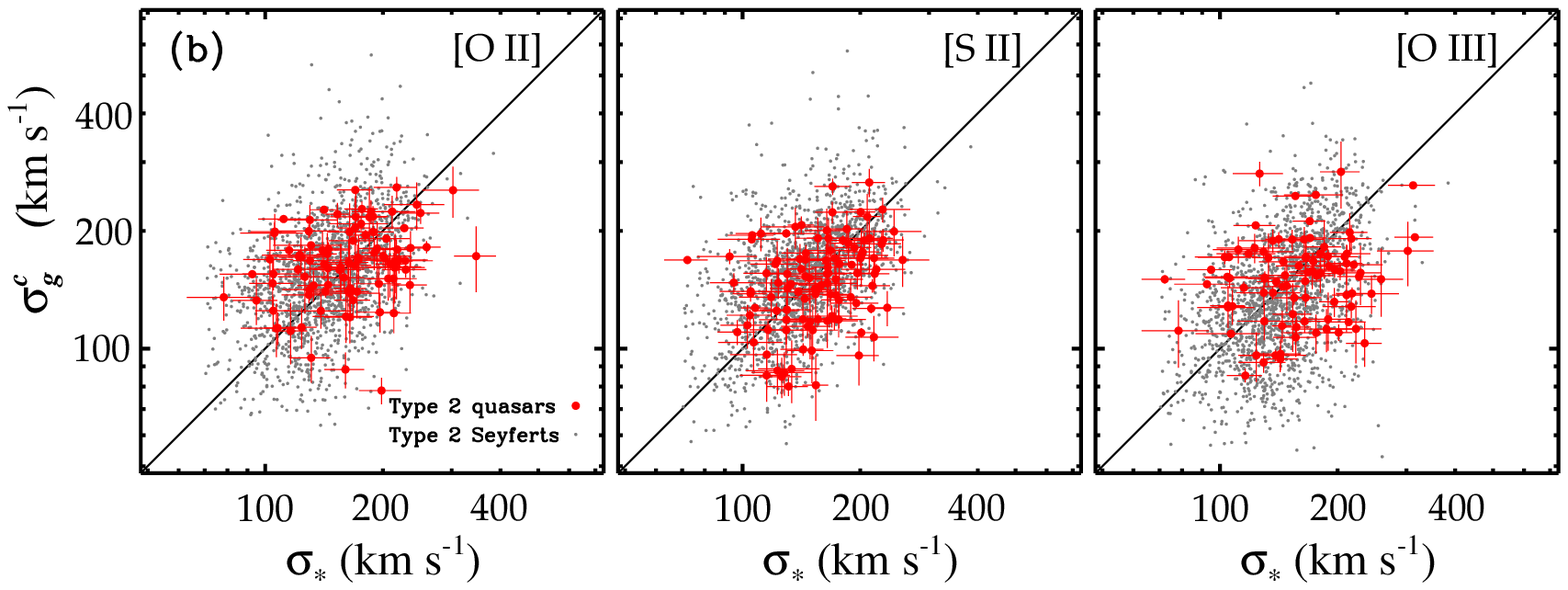}
\vskip -5.0cm
\figcaption[fig9.ps]{
Comparison of stellar velocity dispersions \sigs\ with gaseous velocity 
dispersions for the (left) \oii, (middle) \sii, and (right) \oiii\ emission 
lines measured from (a) the second moment (\siggs) and (b) the
core (\siggc) of the line profile.  The small grey points are from 
the lower luminosity type~2 AGNs (Seyferts) from Greene \& Ho 
(2005a).  The red points with error bars are the subset of type~2 quasars in 
this study with fractional uncertainties $\leq 20\%$.  The solid line denotes 
the 1:1 relation.
\label{fig9}}
\end{figure}

\vskip 0.1cm
\hskip -0.3cm
\begin{figure}
\epsscale{1.}
\plotone{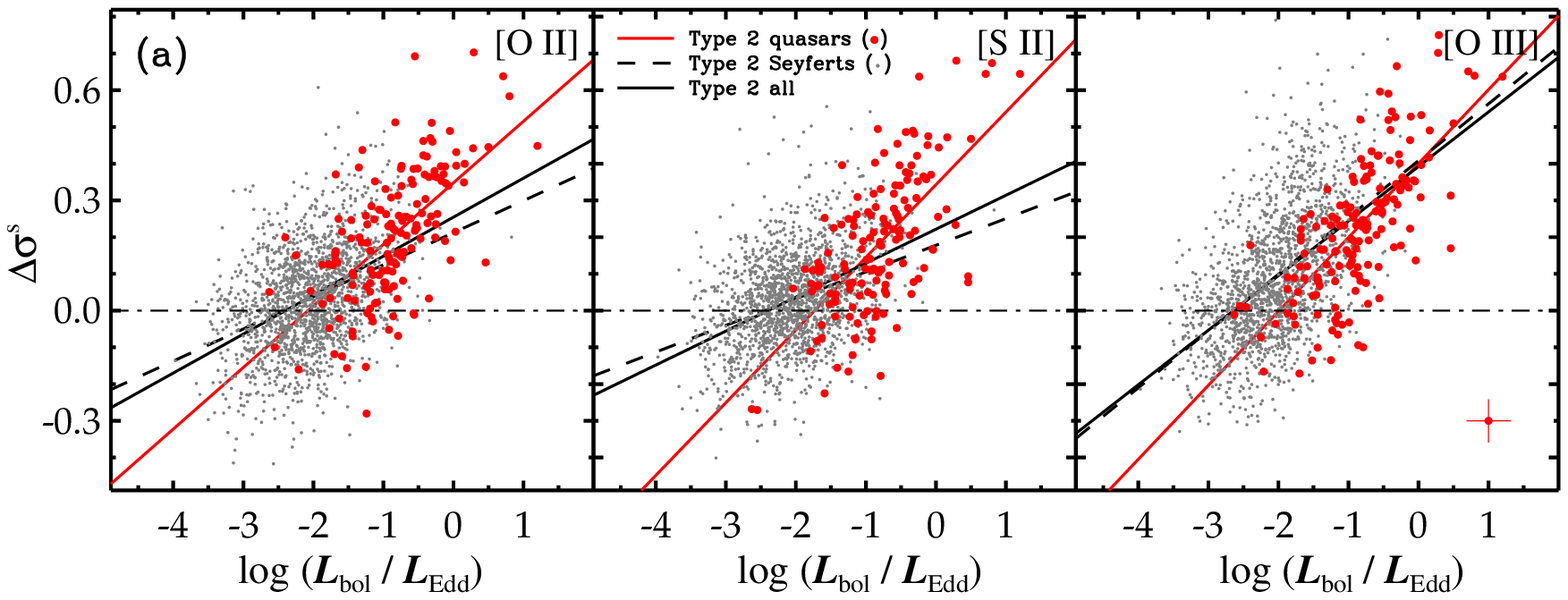}
\vskip -1.0cm
\plotone{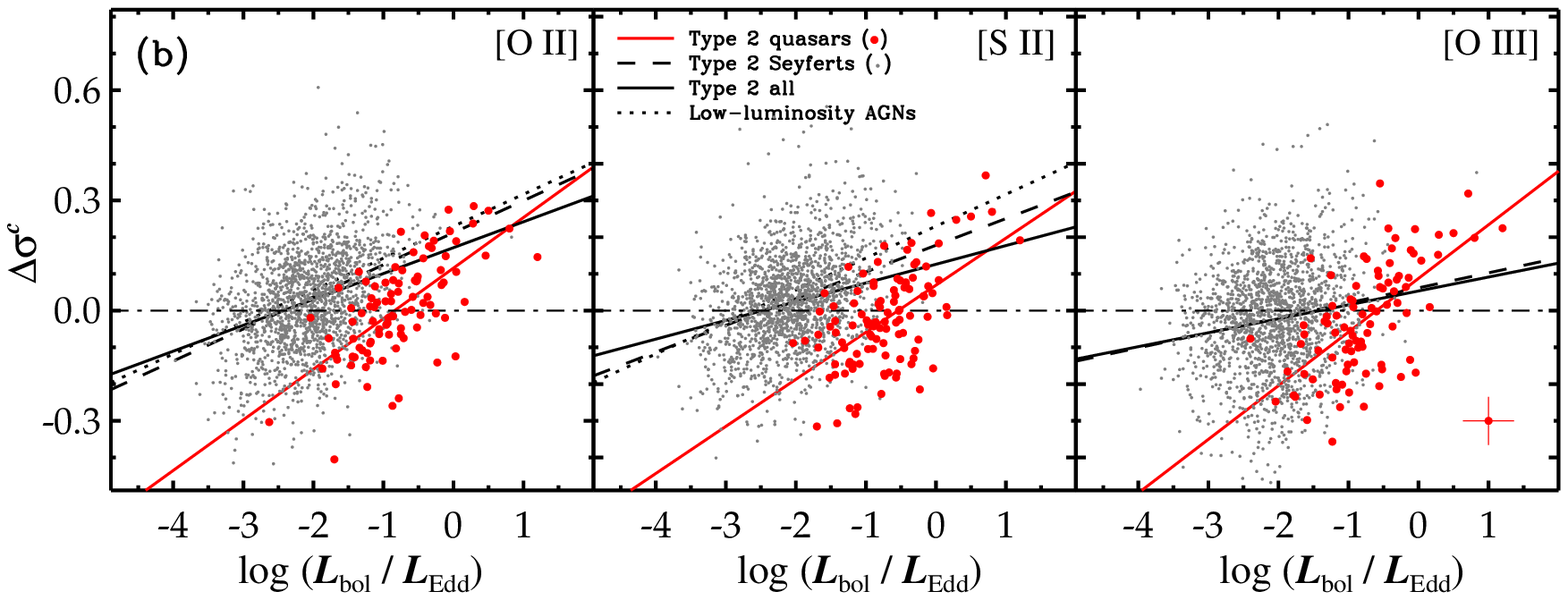}
\figcaption[fig10.ps]{
The relation between \lledd\ and (a) $\Delta\sigma^s \equiv\log\sigma^s_g-
\log\sigma_{\ast}$ and (b) $\Delta\sigma^c \equiv \log\sigma^c_g - 
\log\sigma_{\ast}$ for (left) \oii, (middle) \sii, and (right) \oiii. 
The small grey points are the lower luminosity 
type~2 Seyferts from Greene \& Ho (2005a).  The red points are the subset of 
type~2 quasars in this study with fractional uncertainties $\leq 20\%$; 
individual error bars are omitted for clarity, but a representative error bar 
is shown on the bottom-right corner of the \oiii\ plot.  Best-fit relations, 
calculated using an ordinary least-squares regression with  \lledd\ as the 
independent variable, are given for the type~2 quasars (solid red line), 
type~2 Seyferts (dashed black line), and type~2 Seyferts and quasars combined 
(solid black line). For comparison, the relation derived by Ho (2009; his 
Equation 3) for \nii\ $\lambda 6583$ is overplotted as a dotted black line in 
the \oii\ and \sii\ plots of panel (b). 
\label{fig10}}
\end{figure}

\vskip -10.1cm
\hskip -0.85cm
\begin{figure}
\plotone{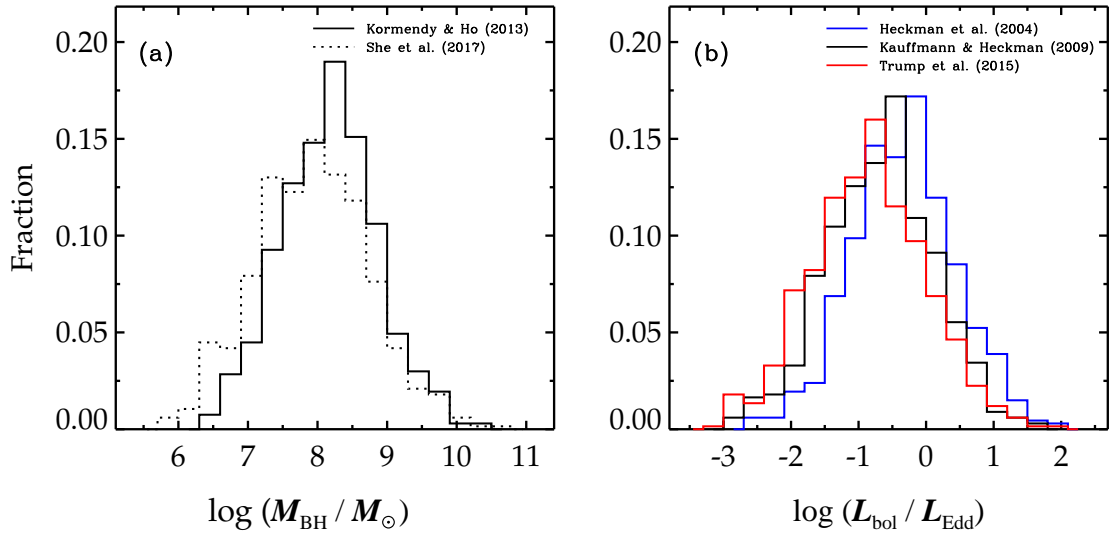}
\figcaption[fig11.ps]{
(a) Distribution of BH masses for our sample of type~2 quasars.  The solid 
histograms assume the \msigma\ relation for classical bulges and elliptical 
galaxies (Equation 2; Kormendy \& Ho 2013); the dotted histograms use the 
\msigma\ relation for all bulge types (see footnote 7; She et al. 2017).  (b) 
Distribution of Eddington ratios assuming the \msigma\ relation for classical 
bulges and elliptical galaxies and the \oiii\ bolometric correction of
Heckman et al. (2004; blue), Kauffmann \& Heckman (2009; black), and
Trump et al. (2015; red), which is based on a power-law fit to the data from 
Lamastra et al. (2009).
\label{fig11}}
\end{figure}

\vskip 0.1cm
\begin{figure}
\vskip .0cm
\epsscale{1.}
\plotone{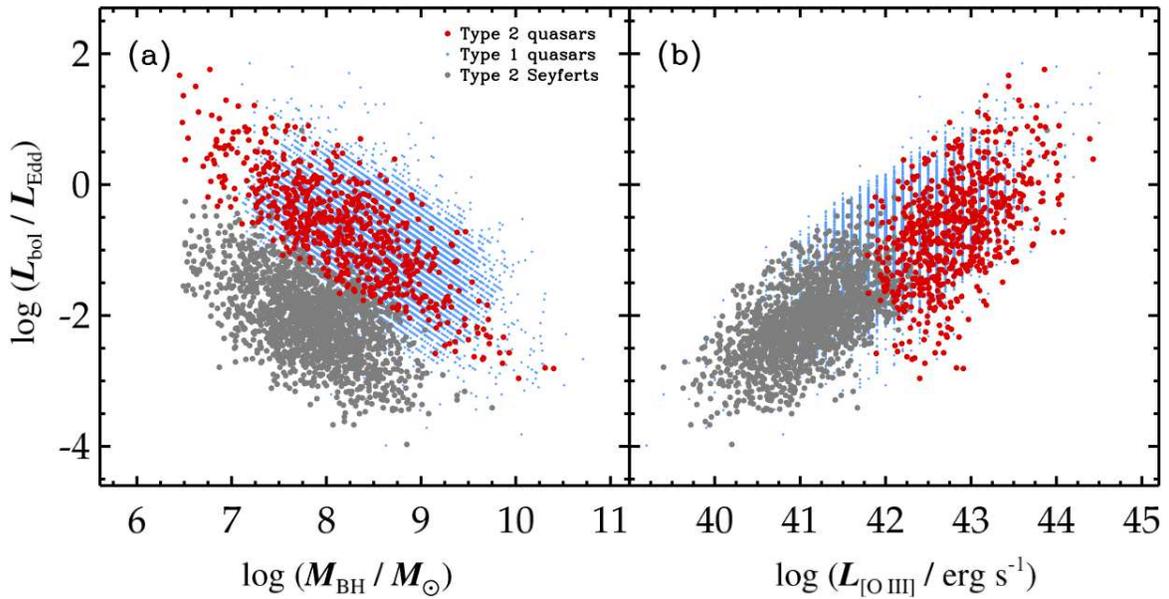}
\figcaption[fig12.ps]{
Distribution of Eddington ratio versus (a) BH mass and (b) \oiii\ luminosity 
for our sample of type~2 quasars (red points), lower luminosity, lower 
redshift type~2 Seyferts from Greene \& Ho (2005a; grey points), and type~1 
quasars with $z < 0.83$ from Shen et al. (2011; blue points).  
\label{fig12}}
\end{figure}


\begin{thebibliography}{}
\bibitem[Abazajian et al. (2009)]{aba09} Abazajian, K., Adelman-McCarthy, J. K., Ag{\"u}eros, M. A., et al. 2009, \apjs, 182, 543
\bibitem[Alexandroff et al. (2013)]{ale13} Alexandroff, R., Strauss, M. A., Greene, J. E., et al. 2013, \mnras, 435, 3306
\bibitem[Antonucci (1993)]{ant93} Antonucci, R. 1993, ARA\&A, 31, 473
\bibitem[Ballantyne (2016)]{bal16} Ballantyne, D. R. 2016, \mnras, 464, 626
\bibitem[Barth et al. (2001)]{bar02}Barth, A. J., Ho, L. C., \& Sargent, W. L. W. 2002, \aj, 124, 2607
\bibitem[Becker \& White (1995)]{beck95}Becker, R. H., White, R. L., \& Helfand, D. J. 1995, \apj, 450, 559
\bibitem[Beifiori et al. (2011)]{beif11}Beifiori, A., Maraston, C., Thomas, D. \& Johansson, J. 2011, \aap, 531, A109
\bibitem[Bender (1990)]{bend90}Bender, R. 1990, A\&A, 229, 441
\bibitem[Bessiere et al. (2012)]{bes12}Bessiere, P., Tadhunter, C. N., Ramos Almeida, C., \& Villar-Martin, M. 2012, \mnras, 426, 276
\bibitem[Bianchi et al. (2017)]{bia17}Bianchi, S., Marinucci, A., Matt, G., et al. 2017, \mnras, 468, 2740
\bibitem[Bruzual et al. (2003)]{bru03}Bruzual, G., \& Charlot, S. 2003, \mnras, 344, 1000
\bibitem[Cappellari \&  Emsellem (2004)]{cap04}Cappellari, M., \& Emsellem, E. 2004, \pasp, 116,138
\bibitem[Cardelli et al. (1989)]{car89}Cardelli, J. A., Clayton, G. C., \& Mathis, J. S., 1989, \apj, 345, 245
\bibitem[Dressler et al. (1984)]{dre84}Dressler, A. 1984, \apj, 286, 97
\bibitem[Du  et al. (2017)]{du17}Du, P., Wang, J.-M., \& Zhang, Z.-X. 2017, \apj, 840, L6
\bibitem[Elitzur \& Ho (2009)]{eli09}Elitzur, M., \& Ho, L. C. 2009, \apj, 701, L91
\bibitem[Elitzur et al. (2014)]{eli14}Elitzur, M., Ho, L. C., \& Trump, J. R. 2014, \mnras, 438, 3340
\bibitem[Elitzur \& Netzer (2016)]{eli16}Elitzur, M., \& Netzer, H. 2016, \mnras, 459, 585
\bibitem[Ferrarese \& Merritt (2000)]{fer00}Ferrarese, L., \& Merritt, D. 2000, ApJ, 539, L9
\bibitem[Filippenko \& Halpern (1984)]{fil84}Filippenko, A. V., \& Halpern, J. P. 1984, \apj, 285, 459
\bibitem[Gallagher et al. (1989)]{gal89}Gallagher, J. S., Bushouse, H., \& Hunter, D. A. 1989, \aj, 97, 700
\bibitem[Ge et al. (2012)]{ge12}Ge, J.-Q., Hu, C., Wang, J.-M., Bai, J.-M., \& Zhang, S. 2012, \apjs, 201, 32
\bibitem[Gebhardt et al. (2000)]{geb00}Gebhardt, K., Bender, R., Bower, G., et al. 2000, ApJ, 539, L13
\bibitem[Gelderman \& Whittle (1994)]{gel94}Gelderman, R., \& Whittle, M. 1994, \apjs, 91, 491
\bibitem[Greene \& Ho (2005a)]{gre05a}Greene, J. E., \& Ho, L. C. 2005a, \apj, 627, 721
\bibitem[Greene \& Ho (2005b)]{gre05b}Greene, J. E., \& Ho, L. C. 2005b, ApJ, 630, 122
\bibitem[Greene \& Ho (2006)]{gre06}Greene, J. E., \& Ho, L. C. 2006, \apj, 641, 117 (GH06)
\bibitem[Greene et al. (2009)]{gre09}Greene, J. E. , Zakamska, N. L., Liu, X., et al. 2009, \apj, 702, 441
\bibitem[Halpern \& Steiner (1983)]{hal83}Halpern, J.~P., \& Steiner, J.~E. 1983, \apj, 269, L37
\bibitem[H\"aring \&Rix (2004) ]{har04}H\"aring, N., \& Rix, H.-W. 2004, \apj, 604, L89
\bibitem[Heckman \& Best (2014)]{hec14}Heckman, T. M., \& Best, P. N. 2014, ARA\&A, 52, 589
\bibitem[Heckman et al. (2004)]{hec04}Heckman, T., Kauffmann, G., Brinchmann, J., et al. 2004, \apj, 613, 109
\bibitem[Ho (2005)]{ho05}Ho, L. C. 2005, \apj, 629, 680
\bibitem[Ho (2008)]{ho08}Ho, L. C. 2008, ARA\&A, 46, 475
\bibitem[Ho (2009)]{ho09}Ho, L. C. 2009, \apj, 699, 638 
\bibitem[Ho et al. (1996)]{ho96}Ho, L. C., Filippenko, A. V., \& Sargent, W. L. W. 1996, \apj, 462, 183
\bibitem[Ho et al. (1997a)]{ho97a}Ho, L. C.,  Filippenko, A. V., Sargent, W. L. W. 1997a, \apjs, 112, 315
\bibitem[Ho et al. (1997b)]{ho97b}Ho, L. C., Filippenko, A. V., Sargent, W. L. W., \& Peng, C. Y. 1997b, \apjs, 112, 391
\bibitem[Ho et al. (2009)]{hol09}Ho, L. C., Greene, J. E., Filippenko, A. V., \& Sargent, W. L. W.  2009, \apjs, 183, 1
\bibitem[Ho \& Kim (2014)]{ho14}Ho, L. C., \& Kim, M. 2014, \apj, 789, 17
\bibitem[Ho \& Kim (2015)]{ho15}Ho, L. C., \& Kim, M. 2015, \apj, 809, 123
\bibitem[Ho et al. (2012)]{ho12}Ho, L. C., Kim, M., \& Terashima, Y. 2012, \apj, 759, L16
\bibitem[Hopkins et al. (2006)]{hop06}Hopkins, P. F., Hernquist, L., Cox, T. J., et al. 2006, \apjs, 163, 1
\bibitem[Kaspi et al. (2000)]{kas00}Kaspi S., Smith P. S., Netzer H., et al. 2000, \apj, 533, 631
\bibitem[Kauffmann \& Heckman (2009)]{kau09}Kauffmann, G., \& Heckman, T. 2009, \mnras, 397, 135
\bibitem[Kelson et al. (2000)]{kel00}Kelson, D. D., Illingworth, G. D., van Dokkum, P. G., \& Franx, M. 2000, ApJ, 531, 159 
\bibitem[Kim et al. (2006)]{kim06}Kim, M., Ho, L. C., \& Im, M. 2006, \apj, 642, 702
\bibitem[Kormendy \& Ho (2013)]{Kor13}Kormendy, J., \& Ho, L. C. 2013, \araa, 51, 511
\bibitem[Lal \& Ho (2010)]{lal10}Lal, D. V., \& Ho, L. C. 2010, \aj, 139, 1089
\bibitem[Lamastra et al. (2009)]{lam09}Lamastra, A., Bianchi, S., Matt, G., et al. 2009, A\&A, 504, 73
\bibitem[Liu et al. (2010)]{liu10}Liu, X., Shen, Y., Strauss, M. A., \& Greene, J. E. 2010, \apj, 708, 427
\bibitem[Liu et al. (2009)]{liu09}Liu, X., Zakamska, N. L., Greene, J. E., et al. 2009, \apj, 702, 1098
\bibitem[Miniutti et al. (2013)]{min13}Miniutti, G., Saxton, R. D., Rodriguez-Pascual, P. M., et al. 2013, \mnras, 433, 1764
\bibitem[Nelson (2000)]{nel00}Nelson, C. H. 2000, \apj, 544, L91
\bibitem[Nelson \& Whittle (1996)]{nel96}Nelson, C. H., \& Whittle, M. 1996, ApJ, 465, 96
\bibitem[Netzer (2015)]{net15}Netzer, H. 2015, ARA\&A, 53, 365
\bibitem[Netzer et al. (2006)]{net06}Netzer, H., Mainier, V., Rosati, P., \& Trakhtenbrot, B. 2006, A\&A, 453, 525
\bibitem[Onken et al. (2004)]{onk04}Onken, C. A., Ferrarese, L., Merritt, D., et al. 2004, \apj, 615, 645
\bibitem[Osterbrock (1989)]{ost89}Osterbrock, D. E. 1989, Astrophysics of Gaseous Nebulae and Active Galactic Nuclei (Mill Valley: Univ. Science Books)
\bibitem[Pennell et al. (2017)]{pen17}Pennell, A., Runnoe, J. C., \& Brotherton, M. 2017, \mnras, 468, 1433
\bibitem[Reyes et al. (2008)]{rey08}Reyes, R., Zakamska, N. L., Strauss, M. A., et al. 2008, \aj, 136, 2373
\bibitem[S$\acute{\rm a}$nchez-Bl$\acute{\rm a}$zquez et al. (2006)]{anc06}S$\acute{\rm a}$nchez-Bl$\acute{\rm a}$zquez, P., Peletier, R. F., Jim$\acute{\rm e}$nez-Vicente, J., et al. 2006, \mnras, 371, 703
\bibitem[Sanders et al. (1988)]{san88}Sanders, D. B., Soifer, B. T., Elias, J. H., et al. 1988, \apj, 325, 74
\bibitem[Sargent et al. (1977)]{sar77}Sargent, W. L. W., Schechter, P. L., Boksenberg, A., \& Shortridge, K. 1977, \apj, 212, 326
\bibitem[Schlafly \& Finkbeiner (2011)]{sch11}Schlafly, E. F., \& Finkbeiner, D. P. 2011, \apj, 737, 103
\bibitem[Schmidt \& Green (1983)]{sch83}Schmidt, M., \& Green, R. F. 1983, \apj, 269, 352
\bibitem[She et al. (2017)]{she17}She, R., Ho, L. C., \& Feng, H. 2017, \apj, 835, 223
\bibitem[Shen et al. (2015)]{shen15}Shen, Y., Greene, J. E., Ho, L. C., et al. 2015, \apj, 805, 96
\bibitem[Shen et al. (2011)]{shen11}Shen, Y., Richards, G. T., Strauss, M. A., et al. 2011, \apjs, 194, 45
\bibitem[Sheth et al. (2003)]{shet03}Sheth, R. K., Bernardi, M., Schechter, P. L., et al. 2003, \apj, 594, 225
\bibitem[She et al. (2017)]{shu17}Shu, X. W., Wang, T. G., Jiang, N., et al. 2017, \apj, 837, 3
\bibitem[Stern \& Laor (2012)]{ster12}Stern, J., \& Laor, A. 2012, \mnras, 426, 2703
\bibitem[Tonry \& Davis (1979) ]{ton79}Tonry, J., \& Davis, M. 1979, \aj, 84, 1511
\bibitem[Tremaine et al. (2002)]{tre02}Tremaine, S., Gebhardt, K., Bender, R., et al. 2002, \apj, 574, 740
\bibitem[Treu et al. (2007)]{tre07}Treu, T., Woo, J.-H., Malkan, M. A., \& Blandford, R. D. 2007, \apj, 667, 117
\bibitem[Trump et al. (2015)]{tru15}Trump, J. R., Sun, M., Zeimann, G. R., et al. 2015, \apj, 811, 26
\bibitem[Urry \& Padovani (1995)]{urr95}Urry, C. M., \& Padovani, P. 1995, \pasp, 107, 803
\bibitem[Valdes et al. (2004)]{val04}Valdes, F., Gupta, R., Rose, J. A.,  Singh, H. P., \& Bell, D. J. 2004, \apjs, 152, 251
\bibitem[Veilleux (1991)]{}Veilleux, S. 1991, \apjs, 75, 383
\bibitem[Vestergaard \& Peterson (2006)]{ves06}Vestergaard, M., \& Peterson, B. M. 2006, \apj, 641, 689
\bibitem[Whittle (1985)]{whi85}Whittle, M. 1985, \mnras, 213, 1
\bibitem[Whittle (1992)]{whi92}Whittle, M. 1992, \apj, 387, 109
\bibitem[Whittle et al. (1988)]{whi88}Whittle, M., Pedlar, A., Meurs, E. J. A., et al. 1988, \apj, 326, 125
\bibitem[Woo et al. (2006)]{woo06}Woo, J.-H., Treu, T., Malkan, M. A., \& Blandford, R. D. 2006, \apj, 645, 900
\bibitem[Woo et al. (2008)]{woo08}Woo, J.-H., Treu, T., Malkan, M. A., \& Blandford, R. D. 2008, \apj, 681, 925
\bibitem[Zakamska et al. (2008)]{zak08}Zakamska, N. L., Gomez, L., Strauss, M. A., \& Krolik, J. H.  2008, \aj, 136, 1607
\bibitem[Zakamska et al. (2016)]{zak16}Zakamska, N. L., Lampayan, K., Petric, A., et al. 2016, \mnras, 455, 4191
\bibitem[Zakamska et al. (2003)]{zak03}Zakamska, N. L., Strauss, M. A., Krolik, J. H., et al. 2003, \aj, 126, 2125
\bibitem[Zhao et al. (2018)]{zha18}Zhao, D., Ho, L. C., Zhao, Y.-Li., Kim, M., \& Shangguan, J. 2018, in preparation
\end{thebibliography}
\end{document}